\documentclass[a4paper,11pt]{article}
\pdfoutput=1
\usepackage{jcapmod} 
\usepackage[scaled]{helvet}
\usepackage[toc,page]{appendix}
\usepackage{mathtools}
\usepackage{booktabs}
\usepackage[english]{babel}
\usepackage{amsmath,amssymb,amsbsy,amstext, amsthm, simplewick, amsfonts}
\usepackage{hyperref}
\usepackage{graphicx}
\usepackage[small]{caption}
\usepackage{siunitx}
\usepackage{upgreek}
\usepackage{framed}
\usepackage{wrapfig}
\usepackage{multirow}
\usepackage{bbm}
\usepackage[svgnames,dvipsnames,x11names]{xcolor}
\usepackage[utf8x]{inputenc}
\usepackage{selinput}
\usepackage{bm}
\usepackage{float}
\usepackage{geometry}
\usepackage{yfonts}
\usepackage{caption}
\usepackage{subcaption}
\usepackage{sidecap}
\usepackage{longtable}
\usepackage{anyfontsize}
\usepackage{dsfont}
\usepackage{tikz}
\usepackage{relsize}
\usepackage{tcolorbox}
\usepackage{shuffle}
\usetikzlibrary{matrix}
\usepackage{bigstrut}
\usepackage{fnpct}
\usetikzlibrary{arrows.meta}
\usetikzlibrary{decorations.pathreplacing,calligraphy}
\usepackage{tensor}
\usepackage{mleftright}

\usetikzlibrary{decorations.markings, decorations.pathmorphing, decorations.pathreplacing, decorations.shapes}

\usepackage{xcolor}
\usepackage{xparse}

\usepackage{fontawesome5}

\makeatletter
\newcommand{\github}[1]{%
   \href{#1}{\faGithub}%
}
\makeatother

\NewDocumentCommand{\colornucleus}{omme{_^}}{%
  \begingroup\colorlet{currcolor}{.}%
  \IfValueTF{#1}
   {\textcolor[#1]{#2}}
   {\textcolor{#2}}
    {%
     #3
     \IfValueT{#4}{_{\textcolor{currcolor}{#4}}}
     \IfValueT{#5}{^{\textcolor{currcolor}{#5}}}
    }%
  \endgroup
}

\usepackage{array}
\newcolumntype{L}[1]{>{\raggedright\let\newline\\\arraybackslash\hspace{0pt}}m{#1}}
\newcolumntype{C}[1]{>{\centering\let\newline\\\arraybackslash\hspace{0pt}}m{#1}}
\newcolumntype{R}[1]{>{\raggedleft\let\newline\\\arraybackslash\hspace{0pt}}m{#1}}

\usepackage[framemethod=default]{mdframed}
\newmdenv[skipabove=7pt,
skipbelow=7pt,
rightline=true,
leftline=true,
topline=true,
bottomline=true,
backgroundcolor=gray!10,
linecolor=black,
innerleftmargin=5pt,
innerrightmargin=5pt,
innertopmargin=5pt,
innerbottommargin=5pt,
leftmargin=0cm,
rightmargin=0cm,
linewidth=1pt]{eBox}

\definecolor{Red}{RGB}{214, 39, 40}
\definecolor{Blue}{RGB} {31, 119, 180}
\definecolor{Orange}{RGB}{255, 153, 51}
\definecolor{Purple}{RGB}{178, 102, 255}
\definecolor{Green}{RGB}{44, 160, 44}
\definecolor{regal}{RGB}{90,0,120}
\definecolor{darkblue}{rgb}{0.15,0.35,0.55}
\definecolor{reddish}{rgb}{0.65, 0.2, 0.2}
\definecolor{darkgreen}{RGB}{50,150,0}
\definecolor{greyish}{rgb}{.90,.90,.90}
\definecolor{greyish2}{rgb}{.96,.96,.96}
\definecolor{greyish3}{rgb}{.37,.37,.37}
\definecolor{darkblue2}{rgb}{0.3,0.4,0.9}
\definecolor{Blue3}{RGB}{31, 119, 180}

\definecolor{blue3}{RGB}{31, 119, 180}
\definecolor{red3}{RGB}{	214, 39, 40}
\definecolor{orange3}{RGB}{255, 127, 14}
\definecolor{green3}{RGB}{44, 160, 44}
\definecolor{repBlue}{RGB}{31, 119, 180}
\definecolor{repRed}{RGB}{	214, 39, 40}
\definecolor{repGreen}{RGB}{44, 160, 44}

\definecolor{vio}{RGB}{19, 130, 164}
\definecolor{vioo}{RGB}{89, 2, 155}
\newcommand{\Comment}[1]{{}}
\usepackage{empheq}

\usepackage[linktocpage=true]{hyperref}
\hypersetup{
colorlinks=true,
citecolor=darkblue,
linkcolor=reddish,
urlcolor=darkblue,
pdfauthor={},
pdftitle={},
pdfsubject={}
}

\newcommand{\dint}{\text{d}}

\usepackage{colortbl}
\definecolor{lightgreen}{cmyk}{0.2, 0, 0.2, 0.2}
\definecolor{lightgray2}{cmyk}{0.1,0.1,0,0.1}
\definecolor{Red2}{RGB}{214, 39, 40}
\definecolor{Blue2}{RGB} {31, 119, 180}
\definecolor{Orange2}{RGB}{255, 127, 14}
\definecolor{Green2}{RGB}{44, 160, 44}

\makeatletter
\newlength{\apb@width}
\newcommand{\autoparbox}[2][c]{\settowidth{\apb@width}{#2}\parbox[#1]{\apb@width}{#2}}

\makeatother


\def\beq{\begin{equation}}
\def\eeq{\end{equation}}
\def\be{\begin{equation}}
\def\ee{\end{equation}}

\DeclarePairedDelimiter\abs{\lvert}{\rvert}

\allowdisplaybreaks[1]
\newcommand\sqmatrix[2][c]{%
  \fixTABwidth{T}%
  \setbox0=\hbox{$\tabbedCenterstack{#2}$}%
  \setstackgap{L}{\dimexpr\maxTAB@width+\tabbed@gap}%
  \tabbedCenterstack[#1]{#2}%
}

\usetikzlibrary{shapes.misc}

\tikzset{cross/.style={cross out, draw=black, minimum size=2*(#1-\pgflinewidth), inner sep=0pt, outer sep=0pt},
cross/.default={1pt}}

\usetikzlibrary{calc}

\begin{document}


\newgeometry{top=2cm, bottom=2cm, left=2cm, right=2cm}

\begin{titlepage}
\setcounter{page}{1} \baselineskip=15.5pt 
\thispagestyle{empty}

\begin{center}
{\fontsize{18}{18} \bf Strongly Coupled Sectors in Inflation:}\vskip 4pt
{\fontsize{14}{14} \bf Gapless Theories and Unparticles}
\end{center}

\vskip 30pt
\begin{center}
\noindent
{\fontsize{12}{18}\selectfont 
Guilherme L.~Pimentel and Chen Yang}
\end{center}

\begin{center}
\vskip 4pt
\textit{Scuola Normale Superiore and INFN, Piazza dei Cavalieri 7, 56126, Pisa, Italy}
\end{center}

\vspace{0.4cm}
\begin{center}{\bf Abstract}
\end{center}
\noindent
We compute correlation functions of the primordial density perturbations when they couple to a gapless, strongly coupled sector of spectator fields---``unparticles"---during inflation. 
We first derive a four-point function of conformally coupled scalars for all kinematic configurations in de Sitter, which exchanges an unparticle at tree level, by performing direct integration using the Mellin-Barnes method. 
To obtain inflationary bispectra and trispectra, we apply weight-shifting operators to the conformally coupled scalar correlator. 
We show that the correlators solve differential equations determined by the additional symmetries enjoyed by the unparticle propagator. 
Based on these differential equations, we are able to discuss the spinning-unparticle exchanges, focusing on two possible cases where the currents or the stress tensor of unparticles are coupled to inflatons, with the help of spin-raising operators. 
Finally, we study the phenomenology of the resulting shape functions. 
Depending on the value of the unparticle scaling dimension, we classify three characteristic shapes for the inflationary bispectra, including near-equilateral, near-orthogonal, and a novel shape which appears when the scaling dimensions are close to half-integers. More generally, we find that the leading order squeezed limits are insufficient to conclusively determine the detection of a light particle or unparticle. Only the full shapes of bispectra and trispectra can break this degeneracy. 

\end{titlepage}
\restoregeometry

\newpage
\setcounter{tocdepth}{2}
\setcounter{page}{2}

\linespread{0.75}
\tableofcontents
\linespread{1.}

\newpage
\section{Introduction}

Cosmological observations point to the existence of a single scalar adiabatic perturbation. 
The fluctuations of the local scale factor at the hot Big Bang source clustering of light and matter, explaining the appearance of structure and order in the late universe, through gravitational clustering. 
What is the microscopic origin of this scalar field? 
In this paper we begin a systematic exploration of ``strongly coupled" scenarios for the microphysics of the density perturbations.

Under the assumption that the universe went through a phase of cosmic inflation, there are many ways of embedding the scalar perturbations in a more complete theory. 
The effective theory of inflation~\cite{Cheung:2007st,Baumann:2009ds,Lee:2016vti} is similar to that of the non-linear sigma model. 
One way of ultraviolet (UV) completing this theory is to consider a linear sigma model in a broken symmetry phase---this is what happens within the Standard Model for the Higgs. 
The analogous mechanism in early-universe cosmology falls under the umbrella of Single Field Inflation. 
Of course, we can then consider more sophisticated scenarios, where the inflaton is weakly coupled to other particles, like in the electroweak sector of the Standard Model. 
In recent years, this class of scenarios\footnote{In its simplest incarnation, models of spectators weakly coupled to the inflaton are called ``quasi-single field inflation" \cite{Chen:2009we,Chen:2009zp,Chen:2010xka,Baumann:2011nk,Assassi:2012zq,Noumi:2012vr}. It was then emphasized that the states in the UV completion can be determined by precise shapes of non-Gaussianities, thus making the analogy with collider physics quantitative---what is now referred to as ``cosmological collider physics"~\cite{Arkani-Hamed:2015bza}.} have been studied in much detail \cite{Maldacena:2002vr,Chen:2010xka,Arkani-Hamed:2015bza,Arkani-Hamed:2018kmz,Baumann:2019oyu,Baumann:2020dch,Aoki:2023tjm}. 
In this direction, the frontier of research is to do precision calculations, either at tree-level and high-point functions, or at loop level. 
The resulting shapes of primordial non-Gaussianity, so-called cosmological correlators, are now well-understood, from their analytic structure to phenomenological consequences. 

The alternative way of UV completing the non-linear sigma model is to think of the scalar fluctuation as a composite particle, made out of yet more elementary constituents. 
This is what happens in Quantum Chromodynamics (QCD) and in models of compositeness for the Higgs \cite{Weinberg:1975gm,Susskind:1978ms,Kaplan:1983fs,Kaplan:1983sm,Dugan:1984hq,Barbieri:2007bh,Bellazzini:2014yua,Panico:2015jxa,Goertz:2018dyw}, etc. 
Such a scenario has been way less explored in this direction; for a few recent examples in this direction see e.g.~\cite{Kumar:2018jxz,Hubisz:2024xnj}. 
We will consider a simpler situation. 
We couple the curvature perturbations weakly to a separate sector, and enlarge the non-linear sigma model to accommodate this coupling. 
This separate sector is strongly coupled. 
The idea is that this is a first step to understand general features of strongly coupled dynamics in inflation. 
Of course it would be interesting to have benchmark examples where background evolution and the scalar curvature perturbation come from compositeness, but we will not attempt to do that here. 

Within this framework, there is still a large array of possibilities for what ``strongly coupled dynamics" means. 
Our choice will be to parametrize the space of models using the mass of the lightest excitation---the mass gap of the theory, $\text{M}_{\text{Gap}}$. 
In this case, there are three possibilities, based on the ratio of the mass gap to the Hubble scale $H$ during inflation. 
In this paper we will study the simplest case, that of a very light mass gap. 
Nonetheless, it is useful to discuss the three major possibilities: 
\begin{itemize}
    \item $\text{M}_{\text{Gap}} \ll H$: If the theory has a very small mass gap, we could approximate it to be a four-dimensional conformal field theory (CFT$_4$). It will have operators with nontrivial anomalous dimensions---so-called ``unparticles," as well as of course a stress tensor, and possibly conserved currents more generally. This scenario will be the main focus of our paper. It has been studied in \cite{Green:2013rd} (see also the recent \cite{Chakraborty:2025myb}), where many general features were described, mostly on squeezed and collapsed limits. We will review some of them, and compute the full shapes of the corresponding cosmological correlator using various techniques.
    \item $\text{M}_{\text{Gap}} \sim H$. Inflation is a good particle detector for masses below, at and slightly above the Hubble scale. \footnote{In scenarios with a chemical potential~\cite{Flauger:2016idt,Bodas:2020yho}, one can probe energy scales up to the strong coupling scale for the pions, which is $\sim 60H$ in slow-roll inflation.} In this case, many bound states of the strongly coupled sector, with masses of order Hubble, will be produced and propagate for a few Hubble volumes before their power gets diluted. They will interfere in an interesting way, potentially making the resulting imprint on cosmological correlators to be novel, compared to the usual quasi-single field inflation. 
     \item $\text{M}_{\text{Gap}} \gg H$. In this case, it is very likely that the imprint of the strongly coupled sector is degenerate with tree-level exchange of the lightest state of the theory, of mass of the order $\text{M}_{\text{Gap}}$, unless there is a large density of states around $\text{M}_{\text{Gap}}$.
\end{itemize}

In this paper we will focus on the first scenario, the gapless case. 
The resulting correlator will have certain features indicative of the anomalous dimension of the operator being exchanged. 
Our main contribution is to compute the full shape, in particular away from the squeezed limit, and also determining the shapes that arise from exchange of spinning operators. 
Perhaps most importantly, we emphasize that the presence of strongly coupled sectors in the early universe has not been systematically explored, and the proposal of organizing this study through the mass gap is a first, still very broad-brush attempt to do so. 
A detailed discussion of the gapped scenarios will be presented in our forthcoming work. 

\subsection*{Outline}
The Section \ref{sec:setup} establishes the theoretical framework by formulating unparticle physics within the effective field theory of inflation. We also comment on using holography as a tool for exploring strongly coupled sectors in inflation more generally. 
In Section \ref{sec:dS_4pt}, we derive an analytic four-point function for conformally coupled scalars in de Sitter space, using the ``in-in" formalism, incorporating tree-level scalar unparticle mediators with generic scaling dimensions. 
We also establish the governing differential equations encoding full kinematic dependence. 
We implement spin-raising operators to determine the correlator when the exchanged unparticles have spin. 
The Section \ref{sec:inf_corr} bridges the correlators of conformally coupled scalars to massless inflatons, via weight-shifting operators. 
We then show how to induce inflationary three-point function with perturbed de Sitter four-point function. 
Their behavior in the collapsed limit and squeezed limit is discussed. 
Phenomenological consequences are analyzed in Section \ref{sec:comments_pheno}, mainly on the shape functions. 
Section \ref{sec:Outlook} concludes the key findings and presents some future directions. 

The technical supplements are provided in the Appendices. 
In Appendix \ref{app:special_functions}, we present basic mathematical properties of special functions that we used throughout the paper. 
In Appendix \ref{app:MB}, we show how to derive the analytic result in detail using Mellin-Barnes integration. 
Appendix \ref{app:SE} contains the shape function for a spinning unparticle spectra template, namely the stress tensor of scalar unparticles. 

\subsection*{Main Results}
The main results of this paper are highlighted here. 
\begin{itemize}
    \item Equation \eqref{result:4pt} is the analytic form for the four-point function of conformally coupled scalars arising from generic scalar unparticle exchanges. 
    This result provides the foundation from which all other correlators are obtained using spin-raising and weight-shifting operations. 
    \item Equation \eqref{eq:boosts}, \eqref{eq:time_trans} and \eqref{eqn:ODE} display the bootstrap equations for our correlator. 
    \item Sections \ref{sec:trispectra} and \ref{sec:bispectra} show that, when $\Delta\geq3/2$ for trispectra and $\Delta\geq2$ for bispectra, the soft limit behaviors are degenerate with the equilateral non-Gaussianity. 
    This fact requires us to look into the full shape, instead of the collapsed/squeezed limit, to search for unparticles. 
    For $\Delta<2$, the squeezed limit degenerates with the exchange of particles in complementary series. 
    Once again, only a comparison of full shapes can conclusively break this degeneracy.
    \item Figure \ref{fig:compare_massive_unparticle} shows the difference between unparticle exchange and massive scalar exchange. 
    \item Table \ref{tab:peak_positions} summarizes the features in shape function for scaling dimension in different regimes. 
          Figures \ref{fig:shape_equil}, \ref{fig:shape_negequil} and \ref{fig:shape_novel} are samples of typical shapes in different regimes. 
\end{itemize}

\subsection*{Notation and Conventions}
We will choose the metric signature to be $(-,+,+,+)$,  
and use natural units $\hbar=c=1$. 
We will use $\varphi$ for conformally coupled scalars and $\phi$ for massless fields. 
The spatial vectors will be denoted as $\vec{x}$ or with Latin letters as $x^i$, $i=1,2,3$. 
Spacetime indices will be denoted using Greek letters, $\mu=0,1,2,3$. 
The momentum of the $n$-th external leg of a correlation function is denoted as $\vec{k}_n$. 
Its magnitude is $k_n\equiv\abs{\vec{k}_n}$. 
The sum of magnitudes of external momenta is denoted as $k_{ij}\equiv k_i+k_j$. 
The scaling dimension of an operator is denoted as $\Delta$, written in the lower index as $\mathcal{O}_\Delta$. 
We denote the spin of a spin-$l$ operator in the upper index as $\mathcal{O}_\Delta^{(l)}$. 

\newpage
\section{Setup}
\label{sec:setup}
In this section, we briefly discuss a setup that accommodates strongly coupled spectator fields in inflation, of various mass gaps. 
If the theory is strongly coupled and has a very large number of degrees of freedom, then we know that its supersymmetric versions have a weakly coupled dual description that is gravitational and holographic~\cite{Hawking:1982dh,Banados:1997df,Maldacena:1997re,Witten:1998qj,Banados:1998dc,Witten:1998zw,Buchel:2002wf,Buchel:2002kj,Aharony:2002cx,Balasubramanian:2002am,Cai:2002mr,Hertog:2004rz,Ross:2004cb,Alishahiha:2004md,Hertog:2005hu,Balasubramanian:2005bg,Hirayama:2006jn,He:2007ji,Cacciapaglia:2007jq,Cacciapaglia:2008ns,Hutasoit:2009xy,Marolf:2010tg,Maldacena:2010un,Chu:2016pea,Kumar:2018jxz,Csaki:2018kxb,Megias:2019vdb,Megias:2021arn,Penin:2021sry,Mishra:2022fic}. 
In that case we can also describe its holographic dual. 
We make a few comments on these holographic duals, mostly to illustrate that the resulting cosmological correlators will have new features. 
Then we go back to our main focus, that of gapless sectors.

\subsection{Unparticle Physics in Inflation}

The infrared (IR) phases of quantum field theories (QFT) can be generally classified into three possibilities: a theory with a mass gap, a theory with massless particles in the IR or a scale-invariant theory with a continuous spectrum \cite{Rychkov:2016iqz}. 
In particle phenomenology, a hypothetical scale-invariant field is called \textit{unparticle}, introduced in \cite{Banks:1981nn,Georgi:2007ek,Georgi:2007si,Georgi:2008pq,Georgi:2009xq,Grinstein:2008qk,Stephanov:2007ry}. 
It is reasonable to consider the unparticles to be conformal fields since we want to capture general features of this scale-invariant sector. 
Some examples of Beyond the Standard Model (BSM) physics that are conformal or approximately conformal include the second Randall-Sundrum model \cite{Randall:1999vf}, the ``hidden-valley" scenario \cite{Strassler:2006im}, and the conformal window in QCD \cite{Gardi:1998ch}---all of them can be regarded as ``unparticle physics".  

It is possible that in the primordial universe, the UV theory contains some fields that become non-trivially scale-invariant as we probe the Hubble scale during inflation. 
Typically they will become strongly coupled and the operators in the spectrum are composites of the UV fields \cite{Banks:1981nn,Sannino:2008nv}. 
This non-trivial scaling invariant sector can possibly be discovered experimentally in some missing energy distributions, for example in \cite{Fox:2011pm,CMS:2014jvv,Sannino:2009za,Cheung:2007zza,Cheung:2007ap}, and in cosmological observations, discussed in \cite{Kikuchi:2007az,Collins:2008ny,McDonald:2008uh,Green:2013rd,Baumgart:2021ptt}. 

Let's review how to couple operators to the metric within the effective field theory (EFT) of inflation, e.g.~\cite{Cheung:2007st,Lee:2016vti}. 
Inflation introduces a natural foliation of spacetime with a clock, so the time-components of the metric $g^{00}$ become acceptable operators on their own. 
\footnote{Within single-field slow-roll inflation, one can see this by simply evaluating the kinetic term of the inflaton on its background value $\bar{\phi}$, $(\nabla \phi)^2 \to g^{00} \dot{\bar{\phi}}^2$.}
In this unitary gauge, for a general action in an Friedmann–Robertson–Walker (FRW) universe constructed from products of any four-dimensional covariant tensors with free upper 0 indices, in the leading order, we can write out 
\begin{align}
    S = \int \dint^4x \sqrt{-g} \left(\frac{1}{2} M_{\text{pl}}^2 R + M_{\text{pl}}^2 \dot{H} g^{00} - M_{\text{pl}}^2 (3H^2+\dot{H}) + \sum_{n=2}^\infty \frac{M_{n}^4(t)}{n!}(g^{00}+1)^n + \dots\right), 
\end{align}
where the limit of slow-roll inflation will correspond to $M_{n}\rightarrow0$. 
If we reparametrize the time coordinate to introduce general covariance $t\mapsto\tilde{t}=t+\pi(t,\vec{x})$, the metric will transform as 
\begin{align}
    g^{00} \rightarrow g^{00} + 2g^{0\mu}\partial_\mu\pi + g^{\mu\nu}\partial_\mu\pi\partial_\nu\pi. 
\end{align}
In the decoupling limit $M_{\text{pl}}\rightarrow\infty$, the longitudinal mode of the metric $\pi$ is the one with strongest interactions, and we can consider it fluctuating around the background metric, while neglecting tensor fluctuations. 
Evaluated in an unperturbed background, the transformation reduces to 
\begin{align}
    g^{00} \rightarrow -1-2\dot{\pi}+(\partial_\mu\pi)^2. 
    \label{def:transf_g00}
\end{align}
Neglecting the linear term above certain energy \cite{Cheung:2007st}, now we construct the interactions when we couple the unparticles to the Goldstone $\pi$. 
In this work, we will be most interested in the interactions between the Goldstone and the scalar unparticles. 
We will also consider the couplings to the unparticle current and the stress tensor of unparticles. 
Since the unparticles are scaling invariant, the FRW curvature will not affect the conservation of the current, while it is natural to consider the traceless part of the stress tensor. 

At leading order in derivatives and to linear order in $\mathcal{O}_\Delta$, the mixing Lagrangian is 
\begin{align}
    \begin{aligned}
        \mathcal{L}^{(s=0)}_{\pi\mathcal{O}} &\equiv \lambda^{4-\Delta} \mathcal{O}_\Delta (g^{00}+1) + \tilde{\lambda}^{4-\Delta} \mathcal{O}_\Delta (g^{00}+1)^2 \\
        &= \lambda^{4-\Delta} \left(-2\dot{\pi}\mathcal{O}_\Delta + (\partial_\mu\pi)^2\mathcal{O}_\Delta\right) + \tilde{\lambda}^{4-\Delta} \left(4\dot{\pi}^2\mathcal{O}_\Delta - 4\dot{\pi}(\partial_\mu\pi)^2\mathcal{O}_\Delta + (\partial_\mu\pi\partial^\mu\pi)^2\mathcal{O}_\Delta\right), 
    \end{aligned}
    \label{def:eft_scalar_int}
\end{align}
where we used \eqref{def:transf_g00} to introduce $\pi$. 
In unitary gauge, the building blocks for spinning fields are $\sigma^{0\dots0}$ and Lorentz-invariant self-interactions, for example $\sigma^{\mu_1\dots\mu_s}\sigma_{\mu_1\dots\mu_s}$. 
There could also exist contractions with the curvature tensors at higher order in derivatives. 
However, since the Lorentz-invariant self-interactions are invariant under all diffeomorphisms, they will not couple to $\pi$ after the Stückelberg trick. 
The spinning fields $\sigma^{0\dots0}$ transform as 
\begin{align}
    \sigma^{0\dots0} \rightarrow (\delta^0_{\mu_1}+\partial_{\mu_1}\pi) \cdots (\delta^0_{\mu_s}+\partial_{\mu_s}\pi) \sigma^{\mu_1\dots\mu_s}. 
    \label{def:transf_sigma}
\end{align} 
In the following discussions, we always take the decoupling limit so that couplings to metric fluctuations become irrelevant. 
For spin-1 unparticle current $J_\mu\equiv\mathcal{O}^{(1)}_{\mu,\Delta=3}$, the mixing Lagrangian in the leading order is 
\begin{align}
    \mathcal{L}^{(s=1)}_{\pi J} \equiv w_1^3 J^0 (g^{00}+1) + w_2^3 J^0 (g^{00}+1)^2, 
\end{align}
where the $w$'s are parameters. 
Using \eqref{def:transf_g00} and \eqref{def:transf_sigma}, we can expand $\mathcal{L}^{(s=1)}_{\pi J}$ as 
\begin{align}
    \mathcal{L}^{(s=1)}_{\pi J} = \frac{w_1^3}{a(t)^2} \left(2J_i\partial_i\pi - (\partial_i\pi)^2 J_0 - 2\dot{\pi}J_i\partial_i\pi\right) + (3w_1^3+4w_2^3)\dot{\pi}^2 J_0 + \dots, 
\end{align}
where $a(t)$ is the FRW scale factor and we used the constraint $\nabla^\mu J_\mu=0$ at the background level, to replace $\dot{\pi}J_0$ by $J_i\partial_i\pi$. 
For spin-2 traceless stress tensor field $T_{\mu\nu}\equiv\mathcal{O}^{(2)}_{\mu\nu,\Delta=4}$, similar to the above discussion, the leading order interactions between $T_{\mu\nu}$ and the Goldstone $\pi$ will be 
\begin{align}
    \mathcal{L}^{(s=2)}_{\pi T} &= \tilde{w}_1^3 T^{00} (g^{00}+1) + \tilde{w}_2^3 T^{00} (g^{00}+1)^2 + \tilde{w}_3^2 T^{\mu\nu} \delta K_{\mu\nu} + \tilde{w}_4^2 T^{\mu\nu} \delta K_{\mu\nu} (g^{00}+1) \\
    &= \tilde{w}_1^3\big(-2\dot{\pi}T_{00} + a(t)^{-2} (\partial_i\pi)^2 T_{00} + 4a(t)^{-2} \dot{\pi}T_{0i}\partial_i\pi\big) - (5\tilde{w}_1^3-4\tilde{w}_2^3) \dot{\pi}^2T_{00} \nonumber \\
    &+ a(t)^{-4} \left(-\tilde{w}_3^2 T_{ij}\partial_i\partial_j\pi+ 2\tilde{w}_4^2\dot{\pi}(\partial_i\partial_j\pi)T_{ij}\right) + \dots, 
\end{align}
where the $\tilde{w}$'s are parameters and $K_{\mu\nu}$ is the extrinsic curvature. 
It is necessary to include the higher-derivative operators to get the relevant interactions for the spatial components $T_{ij}$. 

We usually describe the initial fluctuations at the beginning of the hot Big Bang using the \textit{curvature perturbation} $\zeta(t_i,\vec{x})$, which relates to the Goldstone $\pi$ as 
\begin{align}
    \zeta = -H \pi + O(\pi^2). 
    \label{def:curvature_fluc}
\end{align}
The dimensionless \textit{power spectrum} of $\zeta$ is 
\begin{align}
    \Delta^2_{\zeta} \equiv \frac{1}{4\pi^2}\left(\frac{H}{f_\pi}\right)^4 = (2.100\pm0.030)\times10^{-9}, 
    \label{def:power_spectrum}
\end{align}
where $f_\pi^4$ is the symmetry breaking scale \cite{Baumann:2011su} and we show the amplitude of scalar fluctuations that best fits the Planck satellite observations \cite{Planck:2018vyg}. 

\subsection{Strongly Coupled Sectors with Holographic Duals}
\label{sec:holography}
It is of course difficult to treat strongly coupled field theories, as there are no general tools to study them. 
Besides going to the lattice, a very useful alternative is to use a weakly coupled dual description, if available. 
For some theories, the dual description is gravitational. 
We are considering a case in which we decouple gravity from the scalar dynamics during inflation, and consider a strongly coupled field theory in a rigid background. 
In that case, we can approximate the inflationary geometry to a fixed de Sitter background, and describe the holographic duals of strongly coupled quantum field theories in de Sitter space, see e.g. the discussion in \cite{Marolf:2010tg,Maldacena:2012xp,Hubisz:2024xnj}.

We consider first the case of a gapless sector. 
This case will be analyzed by different means in the rest of the paper, using symmetries and direct integration. 
Nonetheless, it is useful to contrast it with the gapped cases. 
The gravity dual is a weakly coupled theory in a five-dimensional Anti-de Sitter ($\text{AdS}_5$) geometry, with a four-dimensional de Sitter ($\text{dS}_4$) boundary. 
The masses of the light particles (in units of the $\text{AdS}$ radius) correspond to the dimensions of the unparticles, by the usual $\text{AdS/CFT}$ dictionary. 

\begin{figure}[h]
    \begin{subfigure}{0.5\textwidth}
        \centering
        \includegraphics[width=0.5\linewidth]{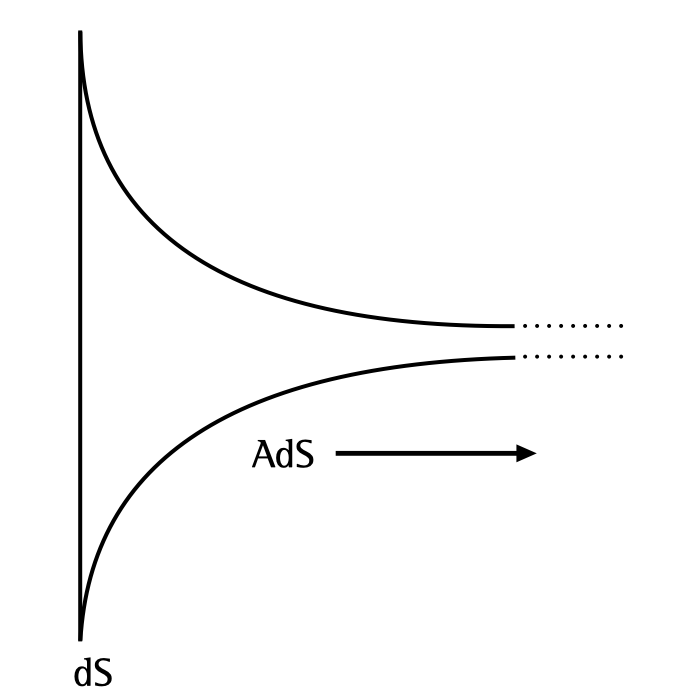}
        \caption{AdS hologram}
        \label{fig:gapless_holo}
    \end{subfigure}
    \begin{subfigure}{0.5\textwidth}
        \centering
        \includegraphics[width=0.5\linewidth]{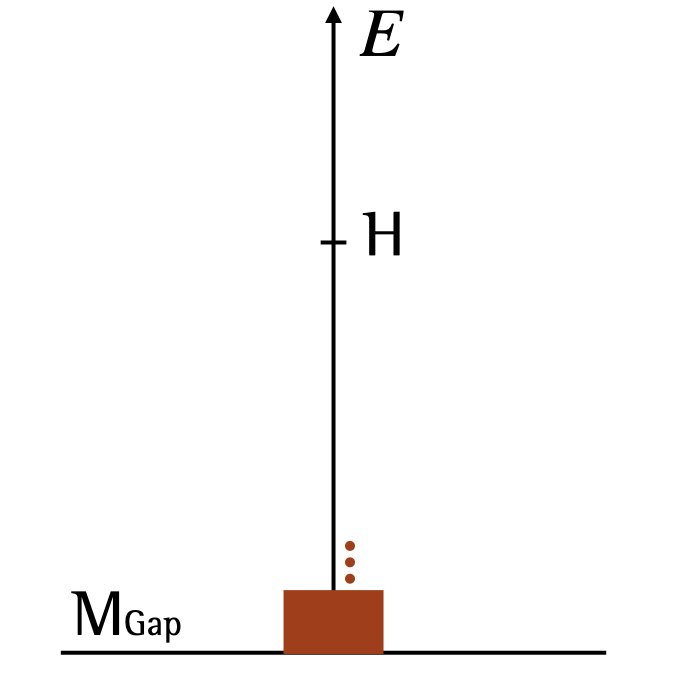}
        \caption{Mass scale compared to Hubble}
        \label{fig:gapless_gap}
    \end{subfigure}
    \caption{Holographic picture for $\text{M}_{\text{Gap}} \ll H$ scenario. The gravity dual of a gapless theory is pure $\text{AdS}$ with a $\text{dS}$ conformal boundary.}
    \label{fig:gapless_ads}
\end{figure}
Now we consider mass deformations of the UV fixed point. 
The resulting gravity duals will depend on the dominant phase of the theory in the large $N$ limit \cite{Hawking:1982dh,Witten:1998zw}. 
We will content ourselves with describing the gravitational duals that correspond to each phase, though at finite $N$ they will both be present and mixed by thermal fluctuations. 
The scenarios are similar to those described in \cite{Randall:1999ee,Randall:1999vf,Maldacena:2010un,Maldacena:2012xp}. 
A useful setup is that of a five-dimensional CFT compactified on $\text{dS}_4 \times S^1$. 
There are two saddles. 
The first one is the Schwarzschild geometry, which when Wick rotated gives a bubble of nothing spacetime \cite{Witten:1998zw,Balasubramanian:2002am,Maldacena:2012xp}, or a spacetime with an end of the world brane. 
This is the gravity dual of a gapped theory with mass much bigger than the Hubble scale of the boundary. 
As the mass gap is rather big, the states are inefficiently produced (being Boltzmann suppressed), so we expect the cosmological correlator in this phase to be given by tree-level exchange of the lightest state of the spectrum---which, being heavy, will be well-approximated by the equilateral shape of non-Gaussianity. 
\begin{figure}[h]
    \begin{subfigure}{0.5\textwidth}
        \centering
        \includegraphics[width=0.65\linewidth]{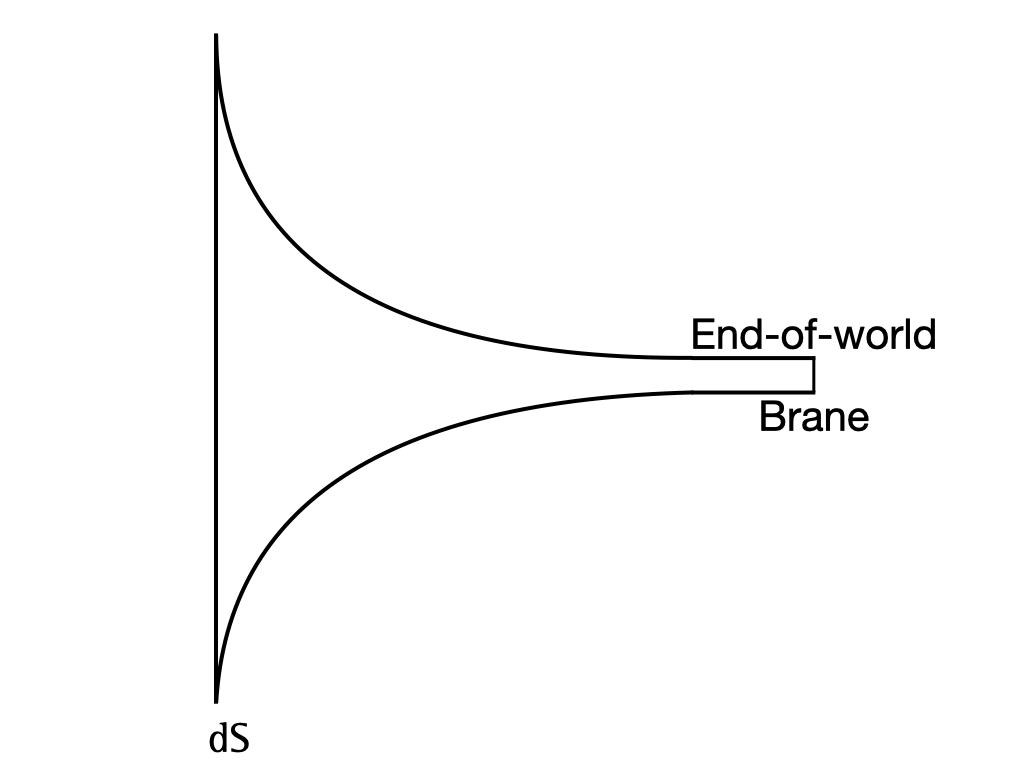}
        \caption{Bubble of nothing hologram}
        \label{fig:cigar_holo}
    \end{subfigure}
    \begin{subfigure}{0.5\textwidth}
        \centering
        \includegraphics[width=0.65\linewidth]{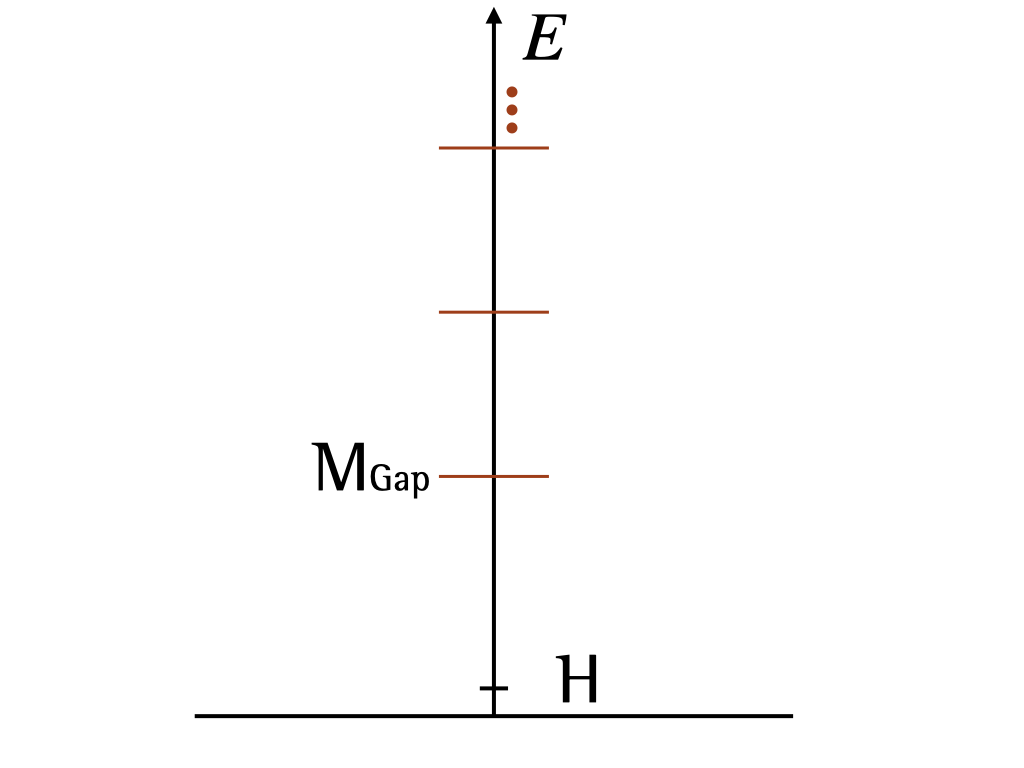}
        \caption{Mass scale compared to Hubble}
        \label{fig:cigar_gap}
    \end{subfigure}
    \caption{Holographic picture for $\text{M}_{\text{Gap}} \gg H$ scenario. The lightest state in the spectrum is heavy in Hubble units, thus the resulting non-Gaussianities in this model will be equilateral-like.}
    \label{fig:cigar}
\end{figure}

Finally, when the mass deformation is not too large, there is a phase of the theory with a gap around the Hubble scale. 
The gravity dual has a holographic renormalization group (RG) flow that can be continued past the location of the putative IR brane, effectively ``frying it."\footnote{Thanks to Raman Sundrum for explaining the appearance of the horizon through ``brane frying."}  
In that case, the dual develops a cosmology behind this brane, a crunching cosmology with open FRW slices---see the discussion in \cite{Maldacena:2012xp}. 
A simple model of this gravity dual is $\text{AdS}_6$ with an identification. 
The identification makes the compactified circle shrink to zero size at the crunch. 

\begin{figure}[h]
    \begin{subfigure}{0.5\textwidth}
        \centering
        \includegraphics[width=0.65\linewidth]{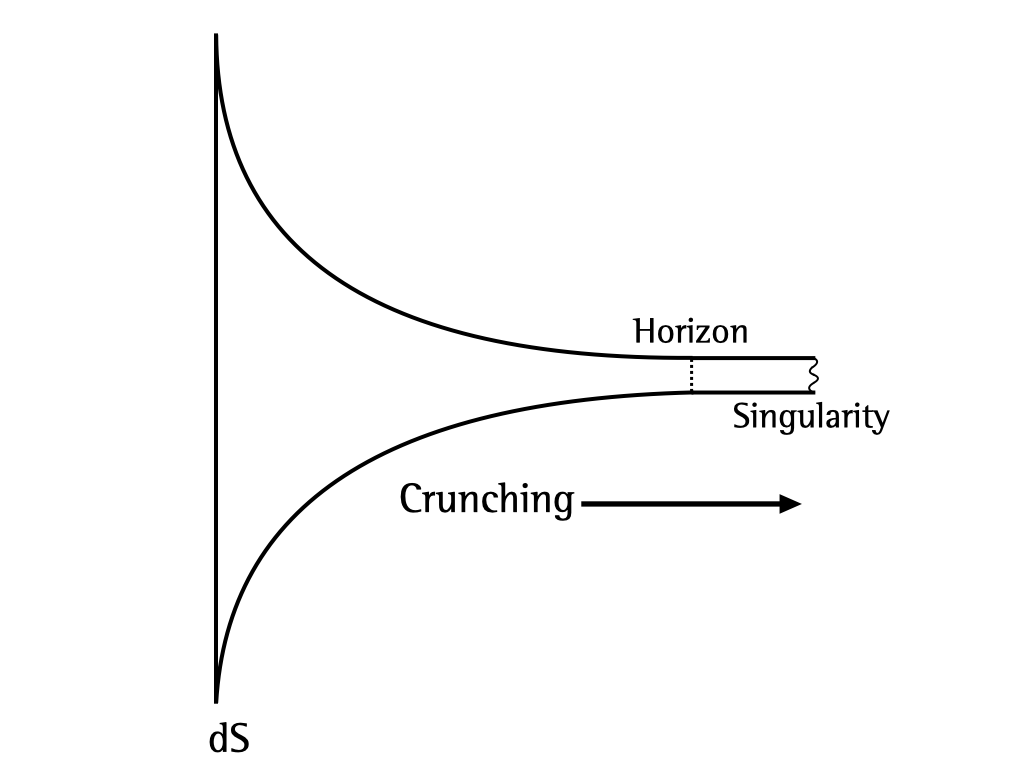}
        \caption{Topological black hole hologram}
        \label{fig:topo_bh_holo}
    \end{subfigure}
    \begin{subfigure}{0.5\textwidth}
        \centering
        \includegraphics[width=0.65\linewidth]{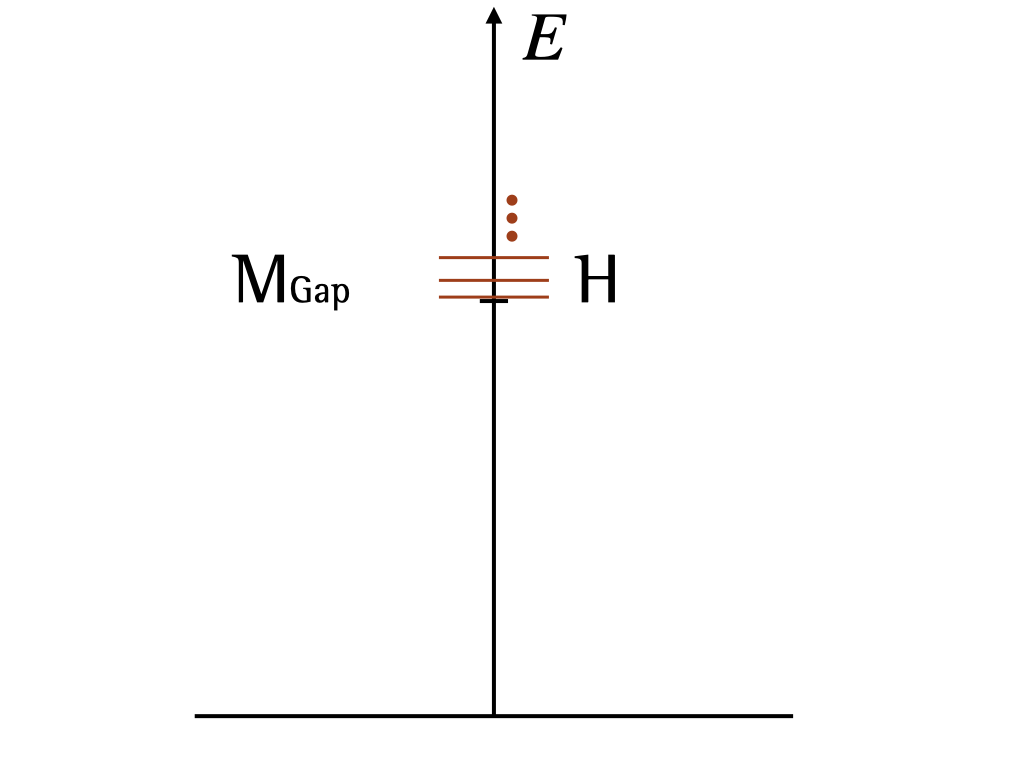}
        \caption{Mass scale compared to Hubble}
        \label{fig:topo_bh_gap}
    \end{subfigure}
    \caption{Holographic picture for $\text{M}_{\text{Gap}} \sim H$ scenario. Many resonances around the Hubble scale will be produced efficiently, interfering with each other and producing a novel shape of non-Gaussianity.}
    \label{fig:topo_bh}
\end{figure}

This is the most exciting case, as we expect many resonances to interfere and produce a non-local, interesting shape of non-Gaussianity. 
In order to determine it, we would study the cosmological correlator of the inflatons exchanging a graviton at tree-level, which is now not trapped in the UV brane, and explores the holographic geometry. 
We show the setup in Figure \ref{fig:topo_bh}. 

We will consider the gapped cases elsewhere. 
For now, we step back and focus on the case of unparticle exchange, for which the gravity dual is not necessary.

\newpage
\section{de Sitter Four-Point Functions}
\label{sec:dS_4pt}

In the following section, we will calculate the de Sitter four-point functions given by scalar unparticle exchange with arbitrary scaling dimension $\Delta$ exactly. 

When computing cosmological correlators, we typically use the flat slicing of de Sitter space, whose metric is given by 
\begin{equation}
    \dint s^2 \equiv h_{\mu\nu}\dint x^\mu \dint x^\nu = \frac{-\dint \eta^2 + \dint \vec{x}^2}{H^2 \eta^2} \equiv \alpha^2(\eta) \dint s^2_{\text{Mink}}, 
    \label{def:dS_metric}
\end{equation}
where $\eta$ is conformal time and $\alpha(\eta)$ is the scale factor. 

From \cite{Arkani-Hamed:2018kmz,Baumann:2019oyu}, we know that we can relate the correlation functions of massless inflatons $\phi$ to the correlation functions of conformally coupled scalars $\varphi$ through weight-shifting operators. 
Thus we consider an effective action of a conformally coupled $\varphi$ coupled to an unparticle field of dimension $\Delta$: 
\begin{align}
    S \equiv S_{\text{CFT}} - \frac{1}{2}\int \dint^4 x \sqrt{-h}\ \left(\partial_\mu \varphi\partial^\mu \varphi + 2H^2 \varphi^2 + \frac{g}{\Lambda^{\Delta-2}} \varphi^2 \mathcal{O}_\Delta\right), 
    \label{def:action_varphi}
\end{align}
where $g$ is the dimensionless coupling constant and $\Lambda$ is the energy scale, below which the unparticle field emerges. 
The mode functions in Bunch-Davies vacuum for the conformally coupled scalars $\varphi$ and the massless fields $\phi$ are 
\begin{align}
    &\hat{f}_{\varphi}(k,\eta) = (-H\eta)\frac{e^{-ik\eta}}{\sqrt{2k}}, \\
    &\hat{f}_{\phi}(k,\eta) = (-H)(1+ik\eta)\frac{e^{-ik\eta}}{\sqrt{2k^3}}. 
\end{align}
Notice that the massless field $\phi$ is related to the fluctuation of the inflaton field, sometimes in the literature it is denoted as $\delta\phi$. 
In the EFT of inflation, this $\phi$ is eaten by the metric, appearing as the Goldstone $\pi$; the conversion is done as usual in unitary gauge. 
The Green's functions with and without time-ordering are conveniently defined based on the mode functions: 
\begin{align}
    \begin{aligned}
            &G_{++}(k;\eta_1,\eta_2) = \hat{f}(k,\eta_1) \hat{f}^*(k,\eta_2) \theta(\eta_1-\eta_2) + \hat{f}^*(k,\eta_1) \hat{f}(k,\eta_2) \theta(\eta_2-\eta_1), \\
            &G_{+-}(k;\eta_1,\eta_2) = \hat{f}^*(k,\eta_1) \hat{f}(k,\eta_2), \\
            &G_{-+}(k;\eta_1,\eta_2) = G_{+-}^*(k;\eta_1,\eta_2), \quad G_{--}(k;\eta_1,\eta_2) = G_{++}^*(k;\eta_1,\eta_2). 
    \end{aligned}
\end{align}

For a conformal field, the scalar two-point function in flat space in Euclidean signature is completely fixed by conformal symmetries, which is 
\begin{equation}
    \langle \mathcal{O}_\Delta (\tau_1,\vec{x}_1)\mathcal{O}_\Delta (\tau_2,\vec{x}_2) \rangle_{\text{flat}} = \frac{1}{\Big((\tau_1 - \tau_2)^2 + (\vec{x}_1-\vec{x}_2)^2\Big)^\Delta}. 
\end{equation}
Here $\Delta$ is the \textit{scaling dimension} of the operator $\mathcal{O}_\Delta$. 

From the scaling relation in \eqref{def:dS_metric}, we can derive the two-point function of conformal fields in de Sitter from the flat space result by taking $\tau\mapsto i\eta$ and the corresponding transformation of operators 
$\mathcal{O}_\Delta(x) \mapsto \tilde{\mathcal{O}}_\Delta(\alpha(\eta) x)\equiv\alpha(\eta)^{-\Delta} \mathcal{O}_\Delta$: 
\begin{align}
    \langle \tilde{\mathcal{O}}_\Delta (\eta_1,\vec{x}) \tilde{\mathcal{O}}_\Delta (\eta_2,\vec{y}) \rangle_{\text{dS}_4} &= \frac{\alpha(\eta_1)^{-\Delta}\alpha(\eta_2)^{-\Delta}}{\Big(-(\eta_1 - \eta_2)^2 + (\vec{x}-\vec{y})^2\Big)^\Delta} = \frac{(H^2\eta_1\eta_2)^\Delta}{\Big(-(\eta_1 - \eta_2)^2 + (\vec{x}-\vec{y})^2\Big)^\Delta}. 
    \label{eqn:2pt_dS4}
\end{align}
In four dimensions, the unitarity bound of scalar conformal fields requires $\Delta\geq1$. 

In inflationary cosmology, since we are computing the equal-time correlator on the late-time boundary, we go to spatial momentum space. From \cite{Grinstein:2008qk,Bautista:2019qxj}, we know that in $d$-dimensional Euclidean space, 
\begin{equation}
    \frac{1}{(2\pi)^{d/2}} \frac{1}{(x^2)^\Delta} = \frac{\Gamma(d/2-\Delta)}{4^{\Delta-d/4}\Gamma(\Delta)} \int \frac{\dint^d k}{(2\pi)^d}\ e^{i\vec{k}\cdot\vec{x}} (k^2)^{\Delta-d/2}. 
    \label{eqn:4d_Fourier}
\end{equation}
Therefore the Fourier transformation of the LHS in $(d-1)$-dimensions will give us 
\begin{align}
    &\int \dint^{d-1} x\ e^{-i\vec{k}\cdot\vec{x}}\frac{1}{(x^2)^\Delta} \nonumber \\
    =\ &\frac{(2\pi)^{d/2} \Gamma(d/2-\Delta)}{4^{\Delta-d/4}\Gamma(\Delta)} \int \frac{\dint k_0}{2\pi}\ e^{ik_0 x^0}\ (k_0^2 + \vec{k}^2)^{\Delta - d/2}, \quad \text{where}\ k^2 \equiv \vec{k}^2 \equiv \abs{\vec{k}}^2, \\
    =\ &\frac{(2\pi)^{d/2}}{4^{\Delta - d/4}\Gamma(\Delta)\sqrt{\pi}} \cdot (2k)^{\Delta - \frac{d-1}{2}} \cdot \frac{1}{\abs{x^0}^{\Delta - \frac{d-1}{2}}} \cdot K_{\Delta - \frac{d-1}{2}}\Big(\abs{x^0} k \Big). 
\end{align}
In four-dimensional de Sitter space, after performing the same transformation in \eqref{eqn:2pt_dS4}, we can see that the time-ordered propagator for unparticles will be 
\begin{align}
    &G_{++}(k;\eta_1,\eta_2) = \frac{H^{2\Delta}\eta_1^\Delta \eta_2^\Delta}{4^{\Delta - 1}\Gamma(\Delta)\sqrt{\pi}} \left(\frac{2k}{i(\eta_1 - \eta_2)}\right)^{\Delta - \frac{3}{2}} K_{\Delta - \frac{3}{2}}\Big(ik(\eta_1 - \eta_2) \Big) \theta(\eta_1 - \eta_2) + c.c\ \theta(\eta_2 - \eta_1), \label{def:deSitter_G_F} \\
    &G_{+-}(k;\eta_1,\eta_2) = \frac{H^{2\Delta}\eta_1^\Delta \eta_2^\Delta}{4^{\Delta - 1}\Gamma(\Delta)\sqrt{\pi}} \left(\frac{2k}{i(\eta_2 - \eta_1)}\right)^{\Delta - \frac{3}{2}} K_{\Delta - \frac{3}{2}}\Big(ik(\eta_2 - \eta_1) \Big), 
\end{align}
after normalizing this expression by $(2\pi)^2$.
If we take $\Delta=1$, \eqref{def:deSitter_G_F} will reduce to the well-known propagator of conformally coupled scalars \cite{Arkani-Hamed:2015bza}: 
\begin{equation}
    G_{++,\Delta=1}(k;\eta_1,\eta_2) = H^2\eta_1 \eta_2 \Big(\frac{e^{-ik(\eta_1-\eta_2)}}{2k}\theta(\eta_1-\eta_2) + \frac{e^{-ik(\eta_2-\eta_1)}}{2k}\theta(\eta_2-\eta_1)\Big). 
\end{equation}

In the ``in-in" formalism (also called \textit{Schwinger-Keldysh} formalism), the four-point correlation function with two-site tree-level exchange in the $s$-channel can always be divided into four sectors \cite{Schwinger:1960qe,Keldysh:1964ud,Jordan:1986ug,Calzetta:1986ey,Maldacena:2002vr,Chen:2010xka,Arkani-Hamed:2015bza}: 
\begin{equation}
    \begin{aligned}
        \langle\varphi_1\varphi_2\varphi_3\varphi_4\rangle &\equiv (2\pi)^3\delta^{(3)} \left(\sum_i\vec{k}_i\right) \langle\varphi_{k_1}\varphi_{k_2}\varphi_{k_3}\varphi_{k_4}\rangle', \\
        \langle\varphi_{k_1}\varphi_{k_2}\varphi_{k_3}\varphi_{k_4}\rangle' &\equiv I_{++} + I_{+-} + I_{-+} + I_{--}.
        \label{def:dS_4pt}
    \end{aligned}
\end{equation}
Since we are working at tree-level, the momentum $k$ of the exchanged unparticle is $k\equiv|\vec{k}_1+\vec{k}_2|$ in the $s-$channel. 
Each sector in \eqref{def:dS_4pt} is defined as 
\begin{equation}
    \begin{aligned}
        & I_{++} = - a_\varphi H^{-2\Delta} \int_{-\infty}^0\int_{-\infty}^0 \frac{\dint\eta_1}{\eta_1^2}\frac{\dint\eta_2}{\eta_2^2}e^{ik_{12}\eta_1}e^{ik_{34}\eta_2}\ G_{++}(k;\eta_1,\eta_2), & &I_{--} = I_{++}^*, \\
        & I_{+-} = a_\varphi H^{-2\Delta} \int_{-\infty}^0\int_{-\infty}^0 \frac{\dint\eta_1}{\eta_1^2}\frac{\dint\eta_2}{\eta_2^2}e^{ik_{12}\eta_1}e^{-ik_{34}\eta_2}\ G_{+-}(k;\eta_1,\eta_2), & & I_{-+} = I_{+-}^*,\\
        & a_\varphi \equiv \frac{g^2 2^2\Lambda^{4-2\Delta}H^{2\Delta}\eta_0^4}{16k_1k_2k_3k_4}, 
    \end{aligned}
    \label{def:four-sectors}
\end{equation}
where $\eta_0$ is the late-time cutoff for the bulk-to-boundary propagator of $\varphi$ and $2^2$ is a symmetry factor. 
The appearance of this late-time cutoff indicates that the correlation functions of conformally coupled scalars $\varphi$ are redshifted away when $\eta_0\rightarrow0$. 
However, using the leading term in the correlation function of $\varphi$, we can easily derive the correlation functions of $\phi$. 
Diagrammatically, $\langle\varphi_{k_1}\varphi_{k_2}\varphi_{k_3}\varphi_{k_4}\rangle'$ can be expressed as 
\begin{equation}
    \langle\varphi_{k_1}\varphi_{k_2}\varphi_{k_3}\varphi_{k_4}\rangle' \equiv \raisebox{-0.7cm}{\includegraphics[width=2.5cm]{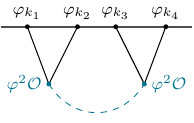}} + \raisebox{-0.7cm}{\includegraphics[width=2.5cm]{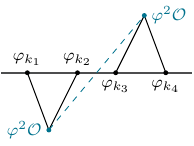}} + \raisebox{-0.7cm}{\includegraphics[width=2.5cm]{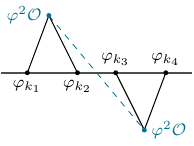}} + \raisebox{-0.3cm}{\includegraphics[width=2.5cm]{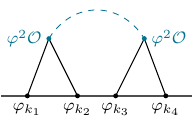}}.
\end{equation}
The full expression of the four-point function requires a sum over permutations. 

\subsection{Mellin-Barnes Integration}
To compute \eqref{def:four-sectors}, it is useful to use the Schwinger parametrization and the integral representation of Bessel function to simplify the integrand: 
\begin{align}
    & \int^\infty_0 {\rm d}x\ e^{ix\eta} x^{a} = (-i\eta)^{-1-a} \Gamma(1+a) \quad \text{when}\ a>-1, \label{def:Sch_para} \\
    & {\rm K}_\nu(z) = \frac{\sqrt{\pi}}{\Gamma(\nu+\frac12)}\big(\frac{z}{2}\big)^{\nu} \int^\infty_1\ e^{-zt}(t^2-1)^{\nu-\frac12}\ {\rm d}t, \quad \text{when } \text{Re}(\nu)>-\frac{1}{2}. \label{def:int_rep_K}
\end{align}
Our convergent region for the integrals will be $\Delta\in(1,2)$. 
$I_{++}$ and $I_{+-}$ can be written as 
\begin{align}
    &\begin{aligned}
        &I_{++} = a_\varphi c_k (-1)^{\Delta}\ \int^\infty_0 \dint x_1 \int^\infty_0 \dint x_2  \int^\infty_1 \dint t \ (x_1 x_2)^{1-\Delta} (t^2-1)^{\Delta-2} \frac{1}{x_1+x_2+X_1+X_2} \nonumber \\
        &\times \left(\frac{1}{x_1+t+X_1} + \frac{1}{x_2+t+X_2} \right), \\
        &I_{+-} = a_\varphi c_k \int^\infty_0 \dint x_1 \int^\infty_0 \dint x_2  \int^\infty_1 \dint t \ (x_1 x_2)^{1-\Delta} (t^2-1)^{\Delta-2} \left(\frac{1}{x_1+t+X_1}\frac{1}{x_2+t+X_2}\right), 
    \end{aligned} \\
    &c_k \equiv \frac{4^{1-\Delta}}{\Gamma^2(2-\Delta)\Gamma(\Delta-1)\Gamma(\Delta)}\frac{1}{k}, 
    \label{def:first_def_I}
\end{align}
where $X_1 \equiv k_{12}/k$, $X_2 \equiv k_{34}/k$. 

From \eqref{def:first_def_I}, it is manifest that the $I_{\pm\pm}$-sectors can be regarded as an integral transform of the corresponding flat-space correlator with shifted ``vertex energies": 
\begin{align}
    I_{\pm\pm,\ \Delta}(X_1,X_2) &\propto \int^\infty_0 \dint x_1 \int^\infty_0 \dint x_2 \ (x_1 x_2)^{1-\Delta} I_{\pm\pm,\ \Delta}^{(\text{flat})} (X_1+x_1,X_2+x_2). 
    \label{eqn:relation_flat}
\end{align}
Furthermore, the flat-space correlator exchanging unparticles is another integral transform of the flat-space correlator exchanging conformally coupled scalars,\footnote{In flat space, conformally coupled fields are massless, but we keep referring them as conformally coupled to avoid confusion with the massless fields in de Sitter.} by sending the ``partial energy" $X+1\mapsto X+t$ and integrating against the kernel 
\begin{align}
    I_{\pm\pm,\ \Delta}^{(\text{flat})}(X_1,X_2) &\propto \int^\infty_1 \dint t \ (t^2-1)^{\Delta-2} I_{\pm\pm,\ \Delta=1}^{(\text{flat})} (X_1,X_2). 
    \label{eqn:relation_cc}
\end{align}
To make the notation consistent with \cite{Arkani-Hamed:2018kmz}, we map $X_1\mapsto1/u$ and $X_2\mapsto1/v$. 
Now the physical regions of the kinematic variables are $u\in(0,1]$, $v\in(0,1]$. 
Based on the derivation explained in detail in Appendix \ref{app:MB}, using Mellin-Barnes integration, we obtain the analytic form of the de Sitter four-point function, given by
\begin{align}
    \begin{aligned}
    &\langle\varphi_{k_1}\varphi_{k_2}\varphi_{k_3}\varphi_{k_4}\rangle' \\
    &= a_\varphi \frac{4^{-\Delta}}{k} \frac{2}{\Delta-1} \Bigg(\Big(\frac{2uv}{u+v}\Big) \frac{1}{2-\Delta} F_1 \Big(1;2-\Delta,2-\Delta;3-\Delta; \frac{u(1-v)}{u+v}, \frac{u(1+v)}{u+v}\Big) \\
    &+ \Big(\frac{2uv}{u+v}\Big) \frac{1}{2-\Delta} F_1 \Big(1;2-\Delta,2-\Delta;3-\Delta; \frac{v(1-u)}{u+v}, \frac{v(1+u)}{u+v}\Big) \\
    &- \Big(\frac{2uv}{u+v}\Big)^{2(\Delta-1)} \frac{1}{\Delta-1} F_1 \Big(1;\Delta-1,\Delta-1;\Delta; \frac{u(1-v)}{u+v}, \frac{u(1+v)}{u+v}\Big) \\
    &- \Big(\frac{2uv}{u+v}\Big)^{2(\Delta-1)} \frac{1}{\Delta-1} F_1 \Big(1;\Delta-1,\Delta-1;\Delta; \frac{v(1-u)}{u+v}, \frac{v(1+u)}{u+v}\Big) \\
    &+ \frac{2}{\Delta-1} F_1\Big(1;\Delta-1,\Delta-1;\Delta;\frac{u-1}{2u},\frac{v-1}{2v}\Big)\Bigg). 
    \end{aligned}    
    \label{result:4pt}
\end{align}
The requirements of integral representations in \eqref{def:Sch_para} and \eqref{def:int_rep_K} make our result \eqref{result:4pt} valid when $\Delta\in(1,2)$. 
To enlarge the valid region for $\Delta$, one can use the analytic properties of gamma functions, which directly relate to the analytic properties of Appell $F_1$ functions, to extend our result to all non-integer $\Delta$'s. 
Some properties of gamma functions and Appell $F_1$ functions are introduced in Appendix \ref{app:special_functions}. 

When $\Delta$ is an integer, there will be UV-divergences in \eqref{result:4pt}. 
Physically, these divergences occurring at integer values of $\Delta$ can be interpreted as the divergences in $n$-loop ``banana" diagrams involving multiple propagators. 
This is explained in detail in \cite{Westerdijk:2025ywh}; the essential idea is that in free field theory one can build integer-valued unparticle operators from taking operator products of free fields. 
This procedure requires proper renormalization, which is reflected in the divergences of the unparticle exchange diagram. 
Moreover, \cite{Bautista:2019qxj} showed that when $\Delta-2\in\mathds{N}^0$, we need to regularize the 2-point function, because the Euclidean 2-point function has no Fourier transform and requires regularization and renormalization, which leads to anomalies. 
The UV-divergences can be absorbed in local counterterms. 
For four-point functions, the counterterms we are able to add to the Lagrangian will be the contact interaction $\lambda\varphi^4$ and its higher-derivatives. 
As we will illustrate with the following examples, these counterterms are able to absorb the UV divergences. 
From \cite{Arkani-Hamed:2018kmz}, we know that contact interactions from irrelevant operators have shapes that can be calculated by repeated application of $\Delta_u$ on the simplest correlator coming from a $\varphi^4$ self-interaction, given by $uv/(u+v)$, with 
\begin{equation}
    \Delta_u \equiv u^2(1-u^2)\partial_u^2 - 2u^3\partial_u. 
\end{equation}

To perform the renormalization for integer $\Delta$'s, one has to expand the Appell $F_1$ functions. 
With the help of the package \textsf{Diogenes} developed in \cite{Bezuglov:2023owj}, we can find $\epsilon$-expansions for Appell $F_1$ functions expanded near integer values. 
The $\Delta=1$ case is well-known \cite{Arkani-Hamed:2015bza,Arkani-Hamed:2018kmz}, it is given by
\begin{align}
    &\langle\varphi_{k_1}\varphi_{k_2}\varphi_{k_3}\varphi_{k_4}\rangle' [\Delta=1]  \nonumber \\
    &= \frac{g^2\Lambda^2H^{2}\eta_0^4}{4k_1k_2k_3k_4k} \left(\text{Li}_2 \Big(\frac{u(1-v)}{u+v}\Big) + \text{Li}_2 \Big(\frac{v(1-u)}{u+v}\Big) + \log\left(\frac{u(1+v)}{u+v}\right)\log\left(\frac{v(1+u)}{u+v}\right) + \frac{\pi^2}{3}\right). 
    \label{eqn:corr_delta_1}
\end{align}
In the $\Delta=2$ case, the diverging part can be isolated with $\epsilon$-expansion of $I_{++}$: 
\begin{align}
    I_{++}[\Delta=2-\epsilon] = \frac{g^2H^{4}\eta_0^4}{4k_1k_2k_3k_4}\frac{1}{4k}\left(- \frac{1}{\epsilon}\frac{uv}{u+v} - \frac{uv}{u+v}\left(\log\left(\frac{u(1+v)}{u+v}\right)+\log\left(\frac{v(1+u)}{u+v}\right)\right)\right). 
\end{align}
One easily sees that the UV-divergent term has the form of a ``total energy" pole, which means it can be absorbed by introducing the contact interaction $\varphi^4$. 
The four-point function after performing renormalization is 
\begin{align}
    &\langle\varphi_{k_1}\varphi_{k_2}\varphi_{k_3}\varphi_{k_4}\rangle' [\Delta=2] \nonumber \\
    &= \frac{g^2H^{4}\eta_0^4}{8k_1k_2k_3k_4k} \left(- \frac{uv}{u+v} \log\left(\frac{(1+u)(1+v)}{u+v}\frac{uv}{u+v}\right) + \frac{uv}{u-v} \log\left(\frac{u(1+v)}{v(1+u)}\right)\right). 
\end{align}
We can perform the same procedure for any integer $\Delta$. 
When $\Delta=3$, we have 
\begin{align}
    &I_{++}[\Delta=3-\epsilon] \nonumber \\
    &= \frac{g^2\Lambda^{-2}H^{6}\eta_0^4}{4k_1k_2k_3k_4}\frac{1}{16k} \Bigg(- \frac{u^2v^2(1+uv)}{(u+v)^3}\log\left(\frac{(1+u)(1+v)}{u+v}\frac{uv}{u+v}\right) - 2\frac{u^2v^2(1+uv)}{(u+v)^3} + \left(\frac{uv}{u+v}\right)^2 \nonumber \\
    &+ \frac{1}{\epsilon} \underbrace{\frac{u^2v^2(1+uv)}{(u+v)^3}}_{\Delta_u\left(\frac{uv}{u+v}\right)}\Bigg). 
\end{align}
The divergent term here is absorbed by a dimension-six contact self-interaction.\footnote{As we are working on the $s$-channel alone, a term with only two derivatives is allowed in the effective action. Of course, if all scalars are identical, it can be removed by a field redefinition.} 
The corresponding four-point function is 
\begin{align}
    &\langle\varphi_{k_1}\varphi_{k_2}\varphi_{k_3}\varphi_{k_4}\rangle' [\Delta=3] \nonumber \\
    &= \frac{g^2\Lambda^{-2}H^{6}\eta_0^4}{32k_1k_2k_3k_4k} \Bigg(- \frac{u^2v^2(1+uv)}{(u+v)^3}\log\left(\frac{(1+u)(1+v)}{u+v}\frac{uv}{u+v}\right) - \frac{2u^2v^2(1+uv)}{(u+v)^3} + \left(\frac{uv}{u+v}\right)^2 \nonumber \\
    &+ \frac{1}{2} \frac{uv}{(u-v)^3}\left(u^2-v^2 - 2uv(u-v) + 2uv(uv-1)\log\left(\frac{u(1+v)}{v(1+u)}\right)\right)\Bigg). 
\end{align}
Again we calculate the $\Delta=4$ result for further usage. 
The sector $I_{++}$ can be expanded as 
\begin{align}
    &I_{++}[\Delta=4-\epsilon] = \frac{g^2\Lambda^{-4}H^{8}\eta_0^4}{4k_1k_2k_3k_4}\frac{4^{-4}}{k} \frac{1}{3} \Bigg(- \frac{8u^3v^3\big(3(1+u^2v^2)+4uv-u^2-v^2\big)}{(u+v)^5}\log\left(\frac{(1+u)(1+v)}{u+v}\frac{uv}{u+v}\right) \nonumber \\
    &- \frac{8u^3v^3\big(8uv-3(1+uv)(u+v)-u^2-v^2+5(1+u^2v^2)\big)}{(u+v)^5} - \frac{2}{\epsilon}\underbrace{\left(-\frac{2u^2v^2(1+uv)}{(u+v)^3}\right)}_{\Delta_u\left(\frac{uv}{u+v}\right)} \nonumber \\
    &+ \frac{1}{\epsilon} \underbrace{\frac{4u^2v^2\big(4uv-u^2-v^2+6u^2v^2(1+uv)-3uv(u^2+v^2)\big)}{(u+v)^5}}_{(\Delta_u)^2\left(\frac{uv}{u+v}\right)}\bigg). 
\end{align}
Still, the UV-divergences are adsorbed by the higher-derivatives of the contact interaction. 
Finally, the four-point function is 
\begin{align}
    &\langle\varphi_{k_1}\varphi_{k_2}\varphi_{k_3}\varphi_{k_4}\rangle' [\Delta=4] \nonumber \\
    &= \frac{g^2\Lambda^{-4}H^{8}\eta_0^4}{192k_1k_2k_3k_4k} \Bigg(- \frac{u^3v^3\big(3(1+u^2v^2)+4uv-u^2-v^2\big)}{(u+v)^5}\log\left(\frac{(1+u)(1+v)}{u+v}\frac{uv}{u+v}\right) \nonumber \\
    &- \frac{u^3v^3\big(8uv-3(1+uv)(u+v)-u^2-v^2+5(1+u^2v^2)\big)}{(u+v)^5} \nonumber \\
    &+ \frac{u^3v^3\big(3(1+u^2v^2)-4uv-u^2-v^2\big)}{(u-v)^5}\log\left(\frac{u(1+v)}{v(1+u)}\right) \nonumber \\
    &+ \frac{uv\left(12u^2v^2(1-uv)(u-v)+\left(6uv(uv-1)+(u-v)^2\right)(u^2-v^2)\right)}{4(u-v)^5}\Bigg). 
\end{align}
After showing the explicit result for all kinematics, we illustrate how it is determined by the additional isometries of the unparticle propagator, and discuss the differential equations satisfied by the four-point function.
\subsection{Differential Equations}
In four dimensions, the isometries of de Sitter space are a subset of the four-dimensional conformal group $SO(4,2)$. 
Since we are considering conformal fields, we can extract extra bootstrap equations from the extra symmetries. 
From \cite{Arkani-Hamed:2018kmz}, the de Sitter isometries requires correlators of conformally coupled scalars exchanging a scalar operator to satisfy 
\begin{equation}
    (\Delta_u-\Delta_v) F(u,v) = 0, 
    \label{def:dS_isometry}
\end{equation}
where $F(u,v)$ is the dimensionless part of the correlator. 
The extra symmetries for the conformal field will be the temporal translational invariance, the boost invariance and the special conformal transformation (SCT) invariance in the time direction. 
However, we can expect that there are only two independent bootstrap equations, since the bulk boost can generate both the temporal translational invariance and temporal SCT from the de Sitter isometries.

\subsubsection*{Temporal Translation}
The temporal translational invariance of \eqref{def:deSitter_G_F} gives us 
\begin{align}
    \Big(\eta_1\eta_2(\partial_{\eta_1} + \partial_{\eta_2}) - \Delta(\eta_1 + \eta_2)\Big) G = 0, 
\end{align}
which imply for $I_{++}$ and $I_{+-}$ the following: 
\begin{align}
    \begin{aligned}
        & \partial_{T,+}I_{++}[\Delta] \equiv \Big((u+v)uv \partial_u \partial_v - (\Delta-1) (u^2\partial_u + v^2\partial_v)\Big) I_{++}[\Delta] = 0, \\
        & \partial_{T,-}I_{+-}[\Delta] \equiv \Big((u-v)uv \partial_u \partial_v - (\Delta-1) (u^2\partial_u - v^2\partial_v)\Big) I_{+-}[\Delta] = 0. 
    \end{aligned}\label{res:tt_inv}
\end{align}

\subsubsection*{Boost}
Now let's turn to the boosts. 
The constraint on the Green's function is 
\begin{align}
    \Big((\eta^2_1 - \eta^2_2)k + (\eta_1\partial_{\eta_1}-\eta_2\partial_{\eta_2})\partial_k \Big) G = 0.
\end{align}
The constraints on $I_{++}$ and $I_{+-}$ from boost will be 
\begin{align}
    \begin{aligned}
        & \partial_{B,+}I_{++}[\Delta] \equiv \Big((u v + u^2 v^2)\partial_u \partial_v + \Delta_u - (\Delta-1)(u\partial_u + v\partial_v - 1)\Big) I_{++}[\Delta] = 0, \\
        & \partial_{B,-}I_{+-}[\Delta] \equiv \Big((u v - u^2 v^2)\partial_u \partial_v + \Delta_u - (\Delta-1)(u\partial_u + v\partial_v - 1)\Big) I_{+-}[\Delta] = 0. 
    \end{aligned}
\end{align}

\subsubsection*{Temporal SCT}
The special conformal transformation in the time direction gives us 
\begin{align}
    \Big((x_1^2+\eta_1^2)\partial_{\eta_1} &+ 2\eta_1 x_1\partial_{x_1} + (x_2^2+\eta_2^2)\partial_{\eta_2} + 2\eta_2 x_2\partial_{x_2} + 2(\eta_1+\eta_2)\Delta\Big)\frac{1}{(H^2\eta_1\eta_2)^\Delta} G = 0.
\end{align}
We can extract a pair of equations for $I_{++}$ and $I_{+-}$, which are obviously just the de Sitter isometry constraint: 
\begin{align}
    \begin{aligned}
        &(\Delta_u-\Delta_v) I_{++}[\Delta] = 0, \quad (\Delta_u-\Delta_v) I_{+-}[\Delta] = 0, 
    \end{aligned}
\end{align}
where we have used the temporal translation and boost invariance during the simplification. 

However, the differential operators $\partial_{T,\pm}$, $\partial_{B,\pm}$ and $(\Delta_u-\Delta_v)$ are not independent. 
They are related by 
\begin{align}
    \begin{aligned}
        \Big(u(1-u^2)\partial_u\partial_{T,\pm} & + v(1\pm uv)\partial_v\partial_{T,\pm} -2\partial_{T,\pm} \\
        &-v(u\pm v)\partial_v\partial_{B,\pm}\Big) I_{+\pm}[\Delta] = -(\Delta-1)u (\Delta_u-\Delta_v) I_{+\pm}[\Delta]. 
    \end{aligned}
\end{align}

Now we are assured that we only expect the correlators exchanging unparticles to satisfy the de Sitter isometry constraint \eqref{def:dS_isometry} and one extra symmetry constraint. 
In order to probe the analytic properties of the correlator, starting from the integral representation \eqref{def:first_def_I} will be more direct for us to derive the differential equation (DE) system controlling our target integral. 

If we consider a general FRW scale factor $\alpha(\eta)\equiv(\eta/\eta_0)^{-(1+\delta)}$ for the metric \eqref{def:dS_metric}, we obtain two ``twist" parameters appearing in the integral representation for the various pieces of the correlator. We denote them as $\psi$ and $\tilde{\psi}$, defined as 
\begin{align}
    \psi &\equiv \int^\infty_0 \dint x_1 \int^\infty_0 \dint x_2  \int^\infty_1 \dint t \ (x_1 x_2)^{\epsilon_1} (t^2-1)^{\epsilon_2}\ \frac{1}{x_1+x_2+X_1+X_2} \left(\frac{1}{x_1+t+X_1}+\frac{1}{x_2+t+X_2}\right), \label{def:general_I++} \\
    \tilde{\psi} &\equiv \int^\infty_0 \dint x_1 \int^\infty_0 \dint x_2  \int^\infty_1 \dint t \ (x_1 x_2)^{\epsilon_1} (t^2-1)^{\epsilon_2}\ \frac{1}{x_1+t+X_1} \frac{1}{x_2+t+X_2}, \label{def:general_I+-}
\end{align}
where the twist parameters are 
\begin{align}
    \epsilon_1 = \delta-(\Delta-1)(\delta+1), \quad \epsilon_2 = \Delta-2. 
\end{align}
We follow the same logic outlined in \cite{Arkani-Hamed:2023kig}, using the techniques of twisted cohomology---in practice, integration-by-parts (IBP) and partial fractioning of the integrand---to derive the differential equation systems obeyed by $\psi$ and $\tilde \psi$. 
First, we define the singular loci of the integrands as 
\begin{align}
    \begin{aligned}
        T_1 &\equiv x_1, & T_3 &\equiv t+1, &B_1 &\equiv x_1+x_2+X_1+X_2, &&\\
        T_2 &\equiv x_2, & T_4 &\equiv t-1, & B_2 &\equiv t+x_1+X_1, \quad B_3 \equiv t+x_2+X_2. 
    \end{aligned}
\end{align}
Now the $\dint\log$-form can be defined using these singularities in order to construct an integral basis for the system: 
\begin{align}
    \dint\log[L_1,L_2,\dots,L_n] \equiv \dint \log(L_1) \wedge \dint \log(L_2) \wedge \dots \wedge \dint \log(L_n), 
\end{align}
where $L_i$'s are any of the $B_i$'s or $T_i$'s. Since we have a triple integral, we pick $\dint\log$'s containing three singularities. 
The $\dint\log$-forms constructed from the singularities will form a complete basis for the differential equation system once we perform the integration together with the twisted part, which is the function with non-integer powers. 

It is obvious that in both \eqref{def:general_I++} and \eqref{def:general_I+-}, the external variables $X_1$ and $X_2$ only come in the combination $x_1+X_1$ and $x_2+X_2$. 
Thus it is not hard to find 
\begin{align}
    &\begin{aligned}
        \dint \psi = -&\epsilon_1 \dint X_1 \int (x_1 x_2)^{\epsilon_1} (t^2-1)^{\epsilon_2} \left(\dint\log[B_1,B_2,T_1] + \dint\log[B_1,B_3,T_1]\right) \\
        +&\epsilon_1 \dint X_2 \int (x_1 x_2)^{\epsilon_1} (t^2-1)^{\epsilon_2} \left(\dint\log[B_1,B_2,T_2] + \dint\log[B_1,B_3,T_2]\right), 
    \end{aligned} \\
    &\begin{aligned}
        \dint \tilde{\psi} &= \epsilon_1 \dint X_1 \int (x_1 x_2)^{\epsilon_1} (t^2-1)^{\epsilon_2} \dint\log[B_2,B_3,T_1] \\
        &+ \epsilon_1 \dint X_2 \int (x_1 x_2)^{\epsilon_1} (t^2-1)^{\epsilon_2} \dint\log[B_2,B_3,T_2]. 
    \end{aligned}
\end{align}

Using the IBP technique described in \cite{Fevola:2024nzj,Tom:2025poss}, the minimal basis of $\dint\log$-forms for the DE system of $\psi$ is 
\begin{align}
    \Omega \equiv 
    \begin{pmatrix}
        &\dint\log[B_1,B_2,T_1] \\
        &\dint\log[B_1,B_2,T_2] \\
        &\dint\log[B_1,B_2,T_3] \\
        &\dint\log[B_1,B_3,T_1] \\
        &\dint\log[B_1,B_3,T_2] \\
        &\dint\log[B_1,B_3,T_3] \\
        &\dint\log[B_1,T_1,T_3]
    \end{pmatrix}, 
\end{align}
and we can further define the vector of integrals, where we add the target integral $\psi$ as the first element: 
\begin{align}
    \vec{I} \equiv \Bigg(\psi,\ \int (x_1 x_2)^{\epsilon_1} (t^2-1)^{\epsilon_2}\ \Omega\Bigg)^{\text{T}}. 
\end{align}
The system of differential equations can be presented in terms of a matrix differential equation
\begin{align}
    \dint \vec{I} = (\epsilon_1 A_1 + \epsilon_2 A_2) \vec{I}, 
\end{align}
with $A_1$ and $A_2$ taking the form 
\begin{align}
    &A_1 = 
    \begin{pmatrix}
        0 & -l_0 & l'_0 & 0 & -l_0 & l'_0 & 0 & 0 \\
        0 & l_1+l_5 & l_1-l_5 & 0 & 0 & 0 & 0 & 0 \\
        0 & l_4-l_5 & l_4+l_5 & 0 & 0 & 0 & 0 & 0 \\
        0 & -l_3 & -l_2 & l_2+l_3 & 0 & 0 & 0 & l_2-l_3 \\
        0 & 0 & 0 & 0 & l_3+l_5 & l_3-l_5 & 0 & 0 \\
        0 & 0 & 0 & 0 & l_2-l_5 & l_2+l_5 & 0 & 0 \\
        0 & 0 & 0 & 0 & -l_1 & -l_4 & l_1+l_4 & l_4-l_1 \\
        0 & 0 & 0 & 0 & 0 & 0 & 0 & 2l_5
    \end{pmatrix}, \\
    &A_2 = 
    \begin{pmatrix}
        0 & 0 & 0 & 0 & 0 & 0 & 0 & 0 \\
        0 & l_1+l_3 & 0 & l_1-l_3 & 0 & 0 & 0 & l_3-l_1 \\
        0 & 0 & l_2+l_4 & l_4-l_2 & 0 & 0 & 0 & l_4-l_2 \\
        0 & 0 & 0 & 0 & 0 & 0 & 0 & 0 \\
        0 & 0 & 0 & 0 & l_1+l_3 & 0 & l_3-l_1 & l_1-l_3 \\
        0 & 0 & 0 & 0 & 0 & l_2+l_4 & l_2-l_4 & l_2-l_4 \\
        0 & 0 & 0 & 0 & 0 & 0 & 0 & 0 \\
        0 & 0 & 0 & 0 & 0 & 0 & 0 & 0
    \end{pmatrix}, 
\end{align}
where we have defined 
\begin{align}
    \begin{aligned}
        &l_0 \equiv \dint X_1, & &l_1 \equiv \dint \log(X_1+1), & &l_3 \equiv \dint \log(X_1-1), && \\
        &l'_0 \equiv \dint X_2, & &l_2 \equiv \dint \log(X_2+1), & &l_4 \equiv \dint \log(X_2-1), & &l_5 \equiv \dint \log(X_1+X_2).
    \end{aligned}
\end{align}
With the same procedure, we can derive the coefficient matrices $\tilde{A}_1$ and $\tilde{A}_2$ for $\tilde{\psi}$, which take simpler forms: 
\begin{align}
    &\tilde{A}_1 = 
    \begin{pmatrix}
        0 & l_0 & l'_0 & 0 \\
        0 & l_1+l'_5 & l_1-l'_5 & 0 \\
        0 & l_2-l'_5 & l_2+l'_5 & 0 \\
        0 & -l_3 & -l_4 & l_3+l_4
    \end{pmatrix}, \\
    &\tilde{A}_2 = 
    \begin{pmatrix}
        0 & 0 & 0 & 0 \\
        0 & l_1+l_3 & 0 & l_1-l_3 \\
        0 & 0 & l_2+l_4 & l_2-l_4 \\
        0 & 0 & 0 & 0
    \end{pmatrix}, 
\end{align}
where we further define $l'_5 \equiv \dint \log(X_1-X_2)$. 

We can extract bootstrap equations for $\psi$ and $\tilde{\psi}$ from these DE systems. 
Up to second order in derivatives, we find 
\begin{align}
    &\left(D_{X_1}-D_{X_2}\right) \psi = 0,\ \left(D_{X_1}-D_{X_2}\right) \tilde{\psi} = 0, \label{eq:boosts} \\
    &\begin{aligned}
        &\Big((X_1+X_2)\partial_{X_1}\partial_{X_2} - \epsilon_1 (\partial_{X_1} + \partial_{X_2})\Big) \psi = 0, \\
        &\Big((X_1-X_2)\partial_{X_1}\partial_{X_2} + \epsilon_1 (\partial_{X_1} - \partial_{X_2})\Big) \tilde{\psi} = 0, 
    \end{aligned}
    \label{eq:time_trans}
\end{align}
where $D_X \equiv (X^2-1)\partial_X^2 - 2(\epsilon_1+\epsilon_2)X\partial_X$. 
From \eqref{eq:boosts} and \eqref{eq:time_trans}, we see that the DEs for $\tilde{\psi}$ relate to the DEs for $\psi$ by taking $X_2\mapsto-X_2$ and multiplying an overall $(-1)$. 
Therefore we will focus on the DE system of $\psi$. 
Taking $X_1\mapsto1/u$, we will find that in de Sitter, where $\delta=0$, \eqref{eq:boosts} becomes our familiar de Sitter isometry equation $\left(\Delta_u-\Delta_v\right) \psi = 0$ and \eqref{eq:time_trans} is the temporal translational invariance \eqref{res:tt_inv} we found from symmetry constraints. 
This ensures the fact that the DE system derived directly from the target integral knows the symmetries of the physical system we want to study. 

We can also extract ordinary differential equations (ODEs) by increasing the order of differentiation. 
Now we will focus on the ODE on $\psi$, since the ODE on $\tilde{\psi}$ can be analyzed in the same procedure. 
From the integral \eqref{def:general_I++}, it is obvious that $\psi$ should be symmetric under $X_1\leftrightarrow X_2$, while the ODEs with respect to $X_1$ and $X_2$ share the same property. 
The differential operator $\text{O}_1$ associated to the $X_1$-ODE is defined as
\begin{align}
    \begin{aligned}
        \text{O}_1 \equiv \partial_{X_1}^2\Bigg(\bigg(&(X_1+X_2)\Big(\big(X_1^2-1\big)\partial_{X_1}^2+2X_1\partial_{X_1}\Big) \\
        &+ (\Delta-1)\Big(2\big(X_1^2-1\big)\partial_{X_1} + 2X_2+\Delta(X_1-X_2)\Big)\bigg)\Bigg). 
    \end{aligned}
    \label{eqn:ODE}
\end{align}
We can see that \eqref{eqn:ODE} reduces to equation (2.33) in \cite{Arkani-Hamed:2018kmz} when $\Delta=1$, in which the exchange is a conformally coupled scalar. 
The differential operator $\text{O}_2$ can be defined easily by shifting $X_1\leftrightarrow X_2$, and we have 
\begin{align}
    \text{O}_1\ \psi\ =\ 0\ =\ \text{O}_2\ \psi. 
    \label{eqn:ord_partial_diff}
\end{align}
We can bootstrap certain solutions of \eqref{eqn:ord_partial_diff} completely by requiring them to have physical singularities. 
A detailed discussion is presented in \cite{Westerdijk:2025ywh}. 

\subsection{Exchange of Spinning Unparticles}

In this section, we will use the scheme developed in \cite{Arkani-Hamed:2018kmz} to derive the de Sitter four-point functions of conformally coupled scalars with spinning unparticle exchange, emphasizing the results about $S=1,\ \Delta=3$ and $S=2,\ \Delta=4$ cases, for conserved currents and the stress tensor of the unparticle. 

In the last section, we have seen that the dimensionless part $F(u,v)$ of the de Sitter four-point correlator satisfies the boost equation 
\begin{equation}
    (\Delta_u - \Delta_v) F(u,v) = 0, 
\end{equation}
which means, based on the discussion in \cite{Arkani-Hamed:2018kmz}, it is valid to apply the same procedure to obtain the correlation functions with higher-spin unparticle exchange. 

To describe the four-point functions with spinning exchanges, we need to introduce more variables: 
\begin{equation}
    \begin{aligned}
        &x = k_1+k_2, & &y = k_3+k_4, & &s = \abs{\vec{k}_1+\vec{k}_2}, \\
        &\alpha = k_1-k_2, & &\beta = k_3-k_4, & &\tau = (\vec{k}_1-\vec{k}_2)\cdot(\vec{k}_3-\vec{k}_4), 
    \end{aligned}
\end{equation}
and we already worked with the dimensionless combinations $u\equiv s/x$ and $v\equiv s/y$. 
It is shown in \cite{Arkani-Hamed:2018kmz} that the four-point function with spin-$S$ can be written as 
\begin{align}
    F_S = \frac{1}{s} \sum_{m=0}^S \bar{\Pi}_m \tilde{\Pi}_{S,m} \hat{A}_{S,m}
    \label{def:spin-ansatz}
\end{align}
where $\bar{\Pi}_m(\tau,\alpha,\beta)$ and $\tilde{\Pi}_{S,m}(\alpha,\beta)$ are the polarization sums of the transverse and longitudinal modes. 
$\bar{\Pi}_m(\tau,\alpha,\beta)$ is defined as 
\begin{align}
    \bar{\Pi}_S(\tau,\alpha,\beta) =& 2^S \sum_{m=0}^{\lfloor S/2\rfloor} C(S,m) \hat{T}^{S-2m} \hat{L}^{2m} \nonumber \\
    =& 2^S \sum_{m=0}^{\lfloor S/2\rfloor} \frac{(-1/2)^m S!}{m!(S-2m)!} \frac{(2S-2m-2)!!}{(2S-2)!!} \left(\frac{s^2\tau+xy\alpha\beta}{s^4}\right)^{S-2m} \nonumber \\
    &\left(\frac{(s^2-x^2)(s^2-y^2)(s^2-\alpha^2)(s^2-\beta^2)}{s^8}\right)^m, 
\end{align}
and for simplicity we can further define 
\begin{align}
    \hat{T} &\equiv \frac{1}{s^2}\left(\tau+\frac{\alpha\beta}{uv}\right), \quad \hat{L}^2 \equiv \left(1-\frac{1}{u^2}\right)\left(1-\frac{1}{v^2}\right)\left(1-\frac{\alpha^2}{s^2}\right)\left(1-\frac{\beta^2}{s^2}\right). 
\end{align}
$\tilde{\Pi}_{S,m}(\alpha,\beta)$ is defined in terms of Legendre polynomials: 
\begin{align}
    \tilde{\Pi}_{S,m}(\alpha,\beta) &\equiv (-1)^m \tilde{P}^m_S (\alpha/s) \tilde{P}^{-m}_S (\beta/s) \nonumber \\
    &= (-1)^m \left(1-\frac{\alpha^2}{s^2}\right)^{-\abs{m}/2} \left(1-\frac{\beta^2}{s^2}\right)^{-\abs{m}/2} P^m_S (\alpha/s) P^{-m}_S (\beta/s). 
\end{align}

The de Sitter isometries constraint on the $u,v$ variables implies that the coefficient function $\hat{A}_{S,m}(u,v)$ satisfies 
\begin{equation}
    \left(\Delta_{m,u} - \Delta_{m,v}\right) \hat{A}_{S,m} = 0, 
\end{equation}
where the differential operator $\Delta_{m,u}$ for each helicity component is 
\begin{align}
    \Delta_{m,u} \equiv u^2(1-u^2)\partial_u^2 - 2u(u^2+m)\partial_u. 
\end{align}
There exists a spin-raising operator, which relates $\Delta_{m,u}$ with different helicities: 
\begin{align}
    D_{uv} \equiv (uv)^2 \partial_u \partial_v \ \Rightarrow\ \Delta_{m,u} D_{uv} = D_{uv}(\Delta_{m-1,u} - 2m). 
\end{align}
This immediately indicates the relation 
\begin{align}
    \hat{A}_{S,m} = D^m_{uv} f_m(u,v)
\end{align}
between the coefficient function $\hat{A}_{S,m}(u,v)$ with helicity $m$ and the corresponding scalar function $f_m(u,v)$ which satisfies $(\Delta_u - \Delta_v) f_m(u,v)=0$. 

Now we have prepared all the tools and we will present explicit results for spin-1 and spin-2 cases, which relate to conserved currents and the stress tensor of the unparticle sector. 

\subsubsection*{Spin-1 Exchange}
The current is the spin-1 operator with lowest scaling dimension in four dimension: $J_\mu\equiv\mathcal{O}_{\mu,\Delta=3}^{(1)}$. 
Based on the ansatz \eqref{def:spin-ansatz}, for spin-1 current exchange we can write\footnote{Strictly speaking, this trispectrum only makes sense for scalar perturbations that carry an additional quantum number, the charge under the unparticle current. Therefore it could only appear in a cross-correlation between curvature and isocurvature perturbations.} 
\begin{align}
    F_J(s,\alpha,\beta,\tau,u,v) = \frac{1}{s} \left(\bar{\Pi}_1 \tilde{\Pi}_{1,1} D_{uv} + \tilde{\Pi}_{1,0} \Delta_u\right) F_{\Delta=3}(u,v). 
    \label{def:spin1_FJ}
\end{align}
The ingredients are 
\begin{align}
    \bar{\Pi}_1 = 2 \hat{T} = \frac{2}{s^2}\left(\tau+\frac{\alpha\beta}{uv}\right), \quad \tilde{\Pi}_{1,1} = \frac{\Gamma(3/2)}{\sqrt{\pi}} = \frac{1}{2}, \quad \tilde{\Pi}_{1,0} = P^0_1\left(\frac{\alpha}{s}\right) P^0_1\left(\frac{\beta}{s}\right) = \frac{\alpha\beta}{s^2}. 
\end{align}
We can plug in our $\Delta=3$ result here: 
\begin{align}
    F_{\Delta=3}(u,v) &= - \frac{u^2v^2(1+uv)}{(u+v)^3}\log\left(\frac{(1+u)(1+v)}{u+v}\frac{uv}{u+v}\right) - \frac{2u^2v^2(1+uv)}{(u+v)^3} + \left(\frac{uv}{u+v}\right)^2 \nonumber \\
    &+ \frac{1}{2} \frac{uv}{(u-v)^3}\left(u^2-v^2 - 2uv(u-v) + 2uv(uv-1)\log\left(\frac{u(1+v)}{v(1+u)}\right)\right). 
\end{align}

\subsubsection*{Spin-2 Exchange}
The spin-2 exchange is also interesting because it is natural to couple the stress-energy tensor to the scalar perturbations. 
We will consider the operator $T_{\mu\nu}=\mathcal{O}_{\mu\nu,\Delta=4}^{(2)}$. 
For spin-2 stress tensor we have 
\begin{align}
    F_T(s,\alpha,\beta,\tau,u,v) = \frac{1}{s} \left(\bar{\Pi}_2 \tilde{\Pi}_{2,2} D_{uv}^2 + \bar{\Pi}_1 \tilde{\Pi}_{2,1} D_{uv} (\Delta_u - 2) + \tilde{\Pi}_{2,0} \Delta_u (\Delta_u - 2) \right) F_{\Delta=4}(u,v). 
    \label{def:spin2_FT}
\end{align}
The ingredients in this case are 
\begin{align}
    &\bar{\Pi}_2 = 4\hat{T}^2-2\hat{L}^2, \quad \tilde{\Pi}_{2,2} = \frac{\Gamma(5/2)}{\sqrt{\pi}2!}=\frac{3}{8}, \quad \tilde{\Pi}_{2,1} = \frac{3\alpha\beta}{2s^2}, \quad \tilde{\Pi}_{2,0} = \frac{(s^2-3\alpha^2)(s^2-3\beta^2)}{4s^4}. 
\end{align}
We can then plug in our $\Delta=4$ result, which is 
\begin{align}
    F_{\Delta=4}(u,v) &= - \frac{u^3v^3\big(3(1+u^2v^2)+4uv-u^2-v^2\big)}{(u+v)^5}\log\left(\frac{(1+u)(1+v)}{u+v}\frac{uv}{u+v}\right) \nonumber \\
    &- \frac{u^3v^3\big(8uv-3(1+uv)(u+v)-u^2-v^2+5(1+u^2v^2)\big)}{(u+v)^5} \nonumber \\
    &+ \frac{u^3v^3\big(3(1+u^2v^2)-4uv-u^2-v^2\big)}{(u-v)^5}\log\left(\frac{u(1+v)}{v(1+u)}\right) \nonumber \\
    &+ \frac{uv\left(12u^2v^2(1-uv)(u-v)+\left(6uv(uv-1)+(u-v)^2\right)(u^2-v^2)\right)}{4(u-v)^5}. 
\end{align}

\newpage
\section{Inflationary Correlators}
\label{sec:inf_corr}
This section starts with weight-shifting operators, which connect the four-point function of conformally coupled scalars $\varphi$ to the corresponding correlation function of massless fields $\phi$. 
Additionally, we will show that by slightly breaking the de Sitter isometries, inflationary three-point function can be induced from its four-point parents, just as in weakly coupled cosmological collider physics. 

\subsection{Weight-Shifting Operators}
Until now, all the correlators that we have obtained are correlators with the external legs to be conformally coupled scalars. 
However the inflaton is nearly massless, so we need to weight-shift the external legs of conformally coupled scalars, as in \cite{Arkani-Hamed:2018kmz,Baumann:2019oyu,Baumann:2020dch,Baumann:2022jpr}. 

To make the relation precise, let us write the \textit{simplified} mode functions of inflatons and conformally coupled scalars as 
\begin{align}
    \phi_k \equiv (1+ik\eta) e^{-ik\eta+i\vec{k}\cdot\vec{x}}, \quad \varphi_k \equiv \eta e^{-ik\eta+i\vec{k}\cdot\vec{x}}. 
\end{align}
Since the scalar perturbation in \eqref{def:eft_scalar_int} couples to the unparticles in a shift-symmetric way, we can directly relate the four-point functions of inflatons to conformally coupled scalars \cite{Arkani-Hamed:2018kmz}, because 
\begin{align}
    \nabla_\mu \phi_{k_1} \nabla^\mu \phi_{k_2} = s^2 U_{12}\left(\varphi_{k_1}\varphi_{k_2}\right), 
\end{align}
where we define the differential operator $U_{12}$ as 
\begin{align}
    U_{12} (\cdot) &\equiv \frac{1}{2} \left(1 - \frac{k_1 k_2}{k_{12}}\partial_{k_{12}}\right) \left(\frac{1-u^2}{u^2}\partial_u (u \cdot) \right). 
\end{align}
From \cite{Arkani-Hamed:2018kmz,Baumann:2019oyu} we know that this implies that the four-point functions $\mathcal{F}(u,v)$ for $\phi$ and $F(u,v)$ for $\varphi$ are related in terms of 
\begin{align}
    \mathcal{F}(u,v) = s^3\ U_{12} U_{34}\ F(u,v). 
\end{align}

This procedure offers us a chance to develop a clear insight into the analytic structure of inflationary correlators. 
However, since $\nabla_\mu \phi_{k_1} \nabla^\mu \phi_{k_2}$ corresponds to $(\partial_\mu\pi)^2$ in the EFT of inflation, the weight-shifting operation also restricts us to certain couplings between the Goldstone $\pi$ and the unparticles $\mathcal{O}_\Delta$. 
A more specific interacting vertex matching requires the de Sitter isometries being more strongly broken, by giving a subluminal speed of propagation to the scalar fluctuations, as discussed in \cite{Pimentel:2022fsc}. 
This is beyond the scope of this work, and we will discuss more in the future directions. 

\subsection{Inflationary Trispectra}
\label{sec:trispectra}
Moving to real inflationary models, we assume a standard slow-roll scenario, where the slow-roll parameter $\epsilon$ which characterizes the deviation from the pure de Sitter is required to be 
\begin{align}
    \epsilon \equiv -\frac{\dot{H}}{H^2} \ll 1. 
\end{align}
With respect to the spirit of \cite{Maldacena:2002vr,Creminelli:2003iq}, the four-point functions as we studied in pure de Sitter at leading order, will not feel this mild symmetry breaking, which makes it possible to calculate the late-time correlator of the curvature perturbation $\zeta$ using the correlator in de Sitter for $\phi$. 

The inflationary \textit{trispectrum} related to the four-point function of inflatons is defined as 
\begin{align}
    \langle\phi_1(\vec{k}_1)\phi_2(\vec{k}_2)\phi_3(\vec{k}_3)\phi_4(\vec{k}_4)\rangle &\equiv \left(\frac{g^2 H^{2\Delta}}{\Lambda^{2\Delta-4}}\right) (2\pi)^3 \delta^{(3)}(\sum_i\vec{k}_i) \prod_i\left(\frac{1}{2k_i^3}\right) \mathcal{T}_{\phi}(k_1,k_2,k_3,k_4). 
\end{align}
The trispectrum can be derived from the four-point correlator of conformally coupled scalars: 
\begin{align}
    \mathcal{T}_{\phi}(k_1,k_2,k_3,k_4) = s^3 \ U_{12} U_{34}\ F(u,v)\ + \ \text{permutations}. 
    \label{def:trispectrum}
\end{align}
To check the consistency of our result with \cite{Green:2013rd}, we consider the behavior in the collapsed limit $s\rightarrow0$: 
in the collapsed limit, the leading contribution from all convergent Appell $F_1$ functions is $1$, which means the scaling behavior is controlled by the term in front of the Appell functions. 
Notice that by properly performing the integration, $\Delta\to3/2$ is not divergent, and we can see the scaling behaviors are different in two regions: 
\begin{enumerate}
    \item when $\Delta<3/2$, $\mathcal{T}_{\phi}$ scales for small $s$ as $s^{2\Delta-3}$; 
    \item when $\Delta\geq3/2$, $\mathcal{T}_{\phi}$ scales for small $s$ as $s^0$. 
\end{enumerate}
Both cases are consistent with the analysis in \cite{Green:2013rd}. 

\subsection{Inflationary Bispectra}
\label{sec:bispectra}
\begin{figure}[h]
    \centering
    \includegraphics[width=0.5\linewidth]{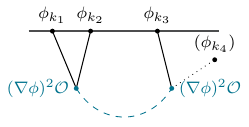}
    \caption{Soft limit of trispectra}
    \label{fig:bispectrum}
\end{figure}

Based on the interacting vertex we considered in \eqref{def:action_varphi}, we cannot construct three-point function with tree-level exchange directly, but we can still induce the inflationary bispectra by evaluating one external leg on the background, as indicated in Figure \ref{fig:bispectrum}. 

However, taking the soft limit $k_4\rightarrow0$ of the four-point function directly will result in zero because in the vertex the inflaton is derivatively coupled. 
Only if we consider a small inflaton mass which is proportional to the slow-roll parameter $\epsilon$, a non-zero result can be obtained. 
The associated mode function will be \cite{Arkani-Hamed:2018kmz}
\begin{align}
    \phi_{k_4,\epsilon} = \left((1+ik_4\eta) + \frac{\epsilon}{2}\log(-k_4\eta)+\dots\right)e^{ik_{4,\mu}x^\mu}. 
    \label{def:mild_broken_inflaton}
\end{align}
Then the four-point function will have a non-trivial soft limit which is proportional to the slow-roll parameter $\epsilon$. 

For the scalar exchange, consider the soft limit for $k_4$ in the operator $\nabla_\mu \phi_{k_3} \nabla^\mu \phi_{k_4,\epsilon}$. 
This operator relates to conformally coupled scalars $\varphi$ as 
\begin{align}
    \lim_{k_4\rightarrow0} \nabla_\mu \phi_{k_3} \nabla^\mu \phi_{k_4,\epsilon} = \frac{\epsilon}{2} k_3^2 \lim_{k_4\rightarrow0}(\varphi_3\varphi_4). 
\end{align}
We denote the dimensionless part of the three-point function of $\varphi$ as $b(u)\equiv F(u,1)$. 
Taking the limit $k_4\rightarrow0$ in \eqref{def:trispectrum}, which means $v\rightarrow1$ from momentum conservation, the induced \textit{bispectrum} with scalar unparticle exchange will be 
\begin{align}
    \mathcal{B}_{\phi}(k_1,k_2,k_3) = \frac{\epsilon}{2}k_3^3 U_{12} b(u) \ +\ \text{permutations}. 
\end{align}
The corresponding three-point function related to $\mathcal{B}_{\phi}$ thus will be 
\begin{align}
    \langle\phi_1(\vec{k}_1)\phi_2(\vec{k}_2)\phi_3(\vec{k}_3)\rangle &\equiv \left(\frac{g^2H^{2\Delta-1}}{\Lambda^{2\Delta-4}}\right) (2\pi)^3 \delta^{(3)}(\sum_i\vec{k}_i) \prod_i\left(\frac{1}{2k_i^3}\right) \mathcal{B}_{\phi}(k_1,k_2,k_3). 
\end{align}
It is always convenient to study the \textit{shape function} of the inflationary three-point function, which is defined as 
\begin{align}
    S(k_1,k_2,k_3) &\equiv \frac{(k_1k_2k_3)^2}{(2\pi)^4\Delta^4_\zeta} \langle\zeta_1(\vec{k}_1)\zeta_2(\vec{k}_2)\zeta_3(\vec{k}_3)\rangle' \nonumber\\
    &\propto (k_1k_2k_3)^2\langle\phi_1(\vec{k}_1)\phi_2(\vec{k}_2)\phi_3(\vec{k}_3)\rangle'. 
\end{align}

The behavior of bispectra in the squeezed limit encodes important features of the underlying dynamics. 
In the squeezed limit, if we take $k_3\rightarrow0$, then $k_1$ will be approximately equal to $k_2$. 
The leading behavior of $S(k_1,k_2,k_3)$ separates into two situations: 
\begin{enumerate}
    \item when $\Delta<2$, $S(k_1,k_2,k_3)$ scales for small $k_3$ as $\left(k_3/k_1\right)^{\Delta-1}$; 
    \item when $\Delta\geq2$, $S(k_1,k_2,k_3)$ scales for small $k_3$ as $\left(k_3/k_1\right)^{1}$. 
\end{enumerate}
The soft behavior for all $\Delta\geq2$ degenerates with the equilateral non-Gaussianity. 
Thus the squeezed limit is not a good detector for unparticles. 
In fact, for some range of complementary series particles, were we to detect them, we'd need more information to break their degeneracy with an unparticle. 

With all these formulas in hand, we can now comment on the phenomenology of the resulting shapes of non-Gaussianity.

\newpage
\section{Comments on Phenomenology}
\label{sec:comments_pheno}
The temperature map of CMB encodes the information of correlations of primordial curvature perturbations, which are proportional to the inflaton fluctuation. 
In this section, we will start by comparing the difference between the four-point functions of a massive scalar exchange and scalar unparticle exchange, and then discuss the similarities and differences between the shapes for inflationary three-point functions with weakly broken conformal symmetry. 
We will also comment on the possibilities to identify this gapless strongly coupled sector in inflation. 
\begin{figure}[ht]
    \centering
    \includegraphics[width=1.0\linewidth]{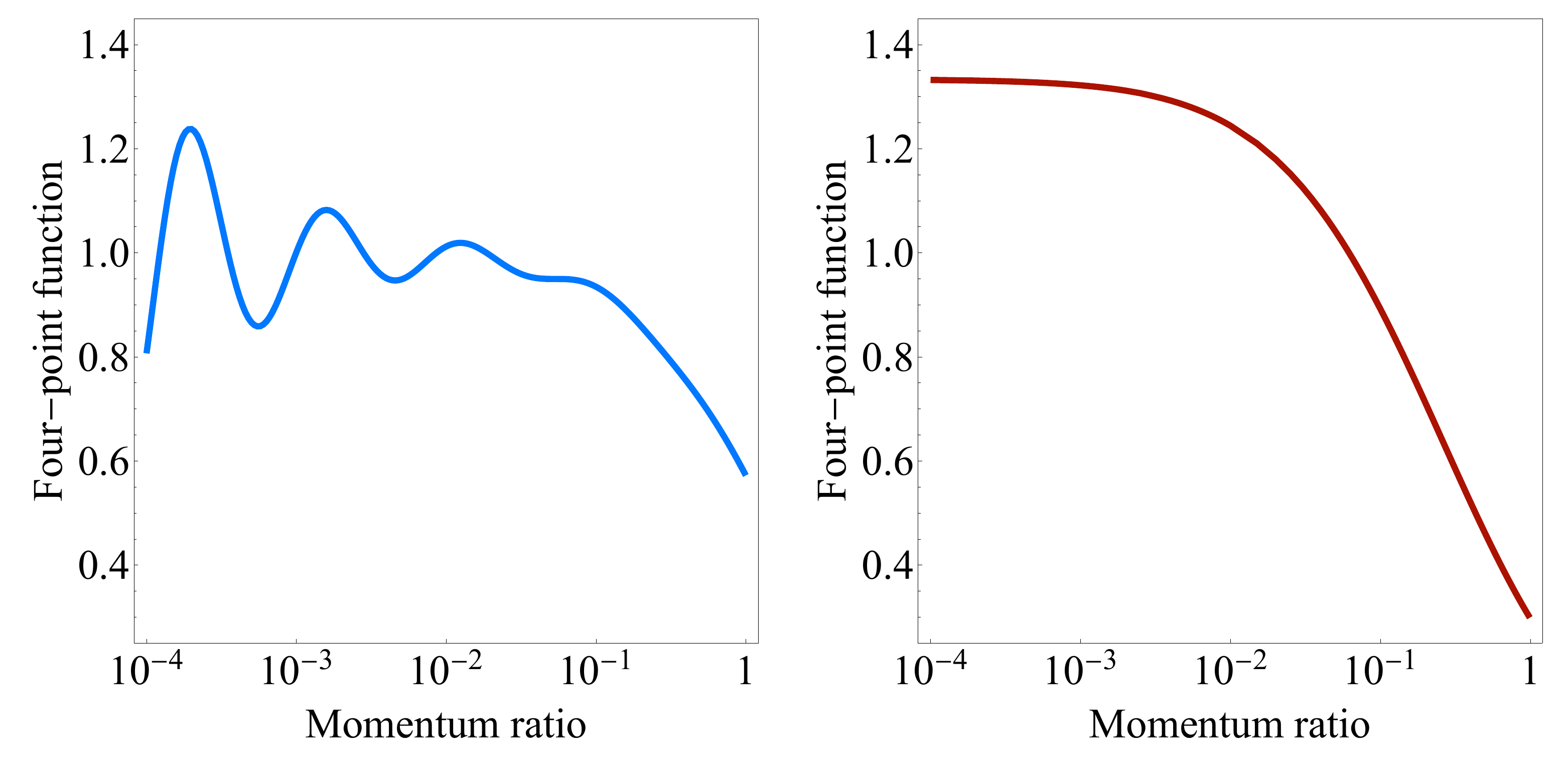}
    \caption{\textit{Left panel: }Example of massive scalar exchange, $u^{-1}\tilde{F}(u,0.5)$, for four-point function of conformally coupled scalars $\varphi$ and an internal particle with $\mu=3$ \cite{Arkani-Hamed:2018kmz}. 
    \textit{Right panel: }Example of scalar unparticle exchange $u^{-1}F(u,0.5)$, for four-point function of $\varphi$ and an internal unparticle with $\Delta=7/2$. 
    Note that we rescaled the four-point functions by $u^{-1}$ in order to visually enhance the squeezed-limit behaviors. To illustrate the shape in the right panel, we also rescale the prefactor in \eqref{result:4pt} which is supposed to significantly suppress the amplitude with the increase of $\Delta$.}
    \label{fig:compare_massive_unparticle}
\end{figure}

Figure \ref{fig:compare_massive_unparticle} reveals a striking dichotomy between massive scalar and unparticle exchanges in the collapsed limit. 
The left panel exhibits characteristic oscillatory modulation for conventional massive scalar exchange. 
In contrast, the four-point function with unparticle exchange (right panel) manifests spectral continuity through the absence of such oscillations. 
This disappearance of oscillatory patterns fundamentally stems from the gapless nature of unparticles. 
The observed phenomenology directly parallels collider physics: where cosmological oscillations correspond to Breit-Wigner resonances in scattering experiments. 
Notably, introducing mass gaps together with strong coupling (at scales $O(\text{M}_\text{Gap})$) would restore resonant features \cite{Strassler:2008bv}. 

\begin{figure}[h]
    \centering
    \includegraphics[width=0.5\linewidth]{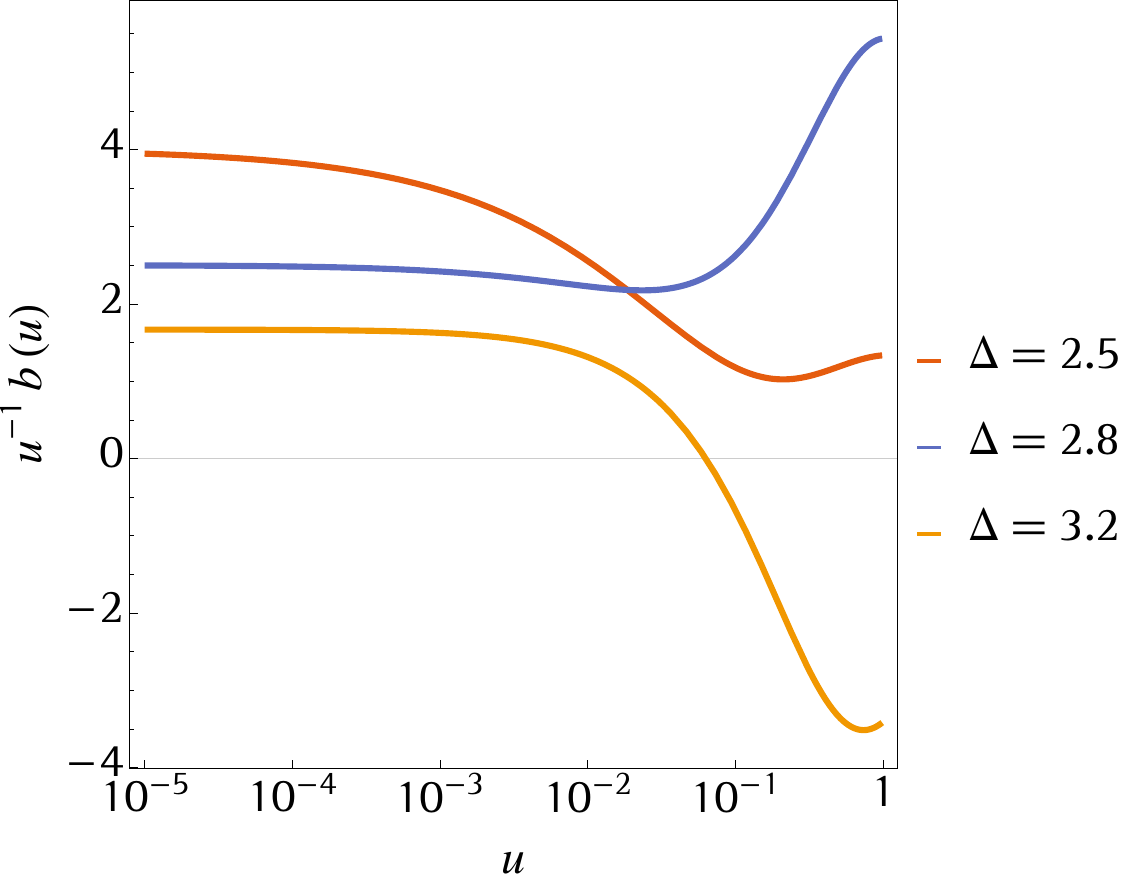}
    \caption{The squeezed limit behavior of different scaling dimensions. 
    When $\Delta>2$, $\lim_{u\rightarrow0} b(u)\sim u$. 
    Note that we also rescaled the three-point functions by $u^{-1}$ in order to visually enhance the squeezed-limit behavior.}
    \label{fig:sq_bispectra}
\end{figure}

The inflationary bispectra are important observables for searching traces of unparticles. 
As discussed in Section \ref{sec:dS_4pt}, the scaling behaviors in the squeezed limit $u\rightarrow0$ of the bispectra can be divided into two classes: when $\Delta<2$, $b(u)$ scales as $u^{\Delta-1}$, and when $\Delta>2$, $b(u)$ scales as $u$ in the squeezed limit. 
In Figure \ref{fig:sq_bispectra}, we show some examples to illustrate similarities in the squeezed-limit behavior of $b(u)$ for $\Delta>2$. 

We will focus on the isosceles triangular shapes ($k_1=k_2=1$), because we are mostly interested in the squeezed limit ($k_3\rightarrow0$) and the equilateral shape ($k_1=k_2=k_3$). 
The maximal peak in different regions of $\Delta$ is summarized in Table \ref{tab:peak_positions}: 
\begin{table}[h]
    \centering
    \begin{tabular}{c|c|c}
        $n\in\mathds{N}_+$ & $1<\Delta<2$ & $\Delta>2$ \\
        \hline
        $n<\Delta<n+1/2$ & Equilateral & (Negatively) Equilateral \\
        \hline
        $\Delta=n+1/2$ & Equilateral & Oscillating when $\Delta>3$ \\
        \hline
        $n+1/2<\Delta<n+1$ & Equilateral & Equilateral \\
    \end{tabular}
    \caption{A summary of the position of peaks in different regions}
    \label{tab:peak_positions}
\end{table}

\begin{figure}[h]
    \centering
    \includegraphics[width=0.42\linewidth]{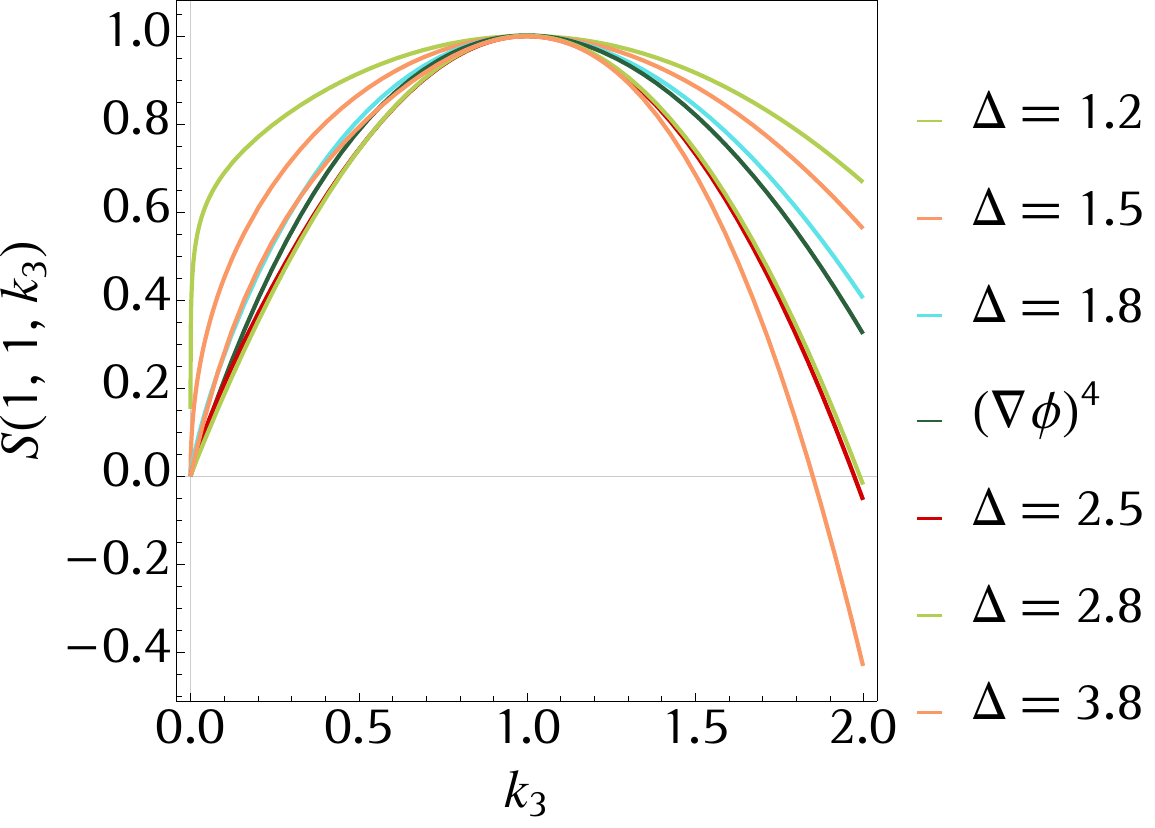}
    \caption{The shape functions of $\Delta$'s that peak at the equilateral shape. Notice that the shapes are rescaled to have the same extrema for comparison.}
    \label{fig:shape_equil}
\end{figure}

\begin{figure}[h]
    \centering
    \includegraphics[width=0.70\linewidth]{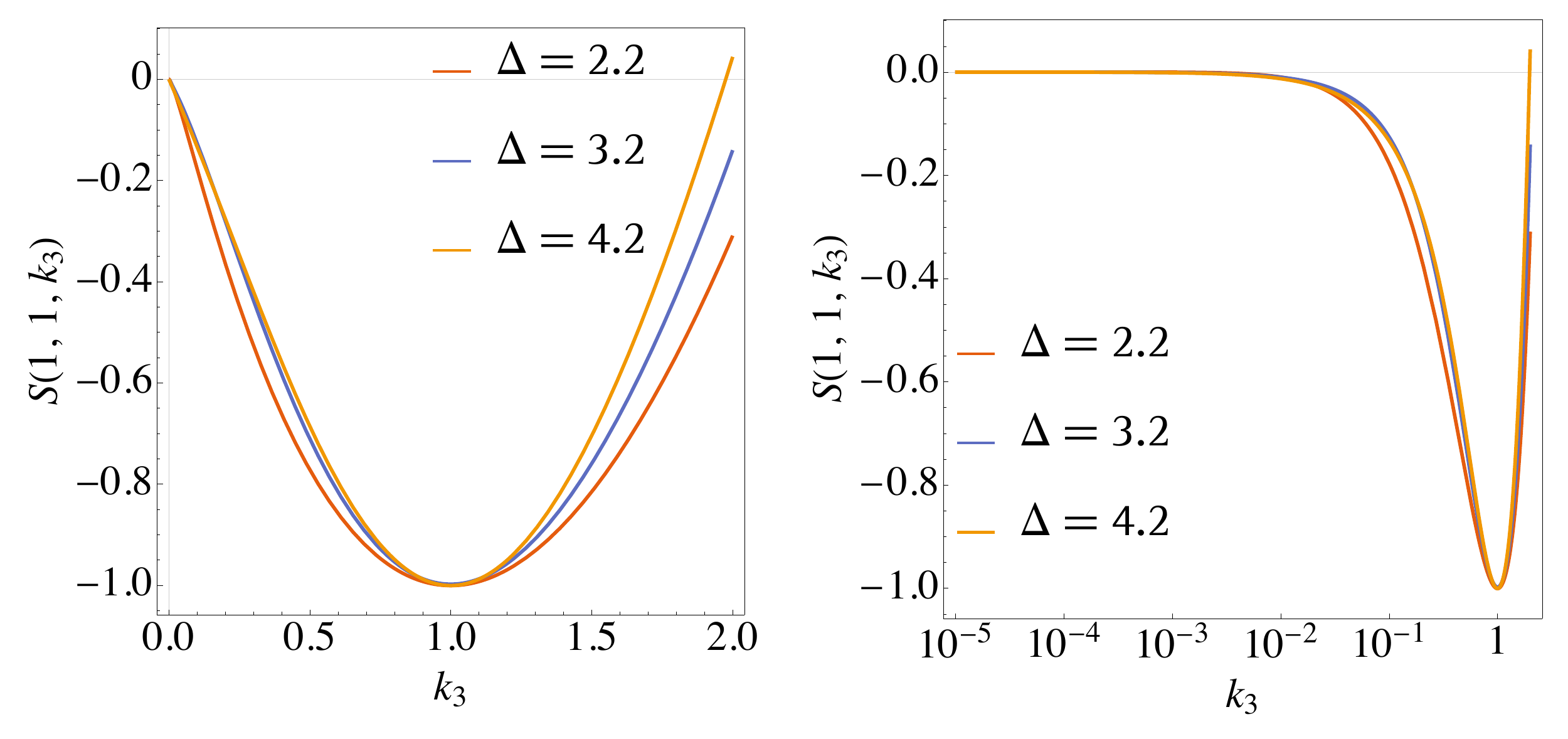}
    \caption{The shape functions of $\Delta$'s that negatively peak at the equilateral shape. We also rescale the magnitudes to have the same value in the equilateral shape for comparison.}
    \label{fig:shape_negequil}
\end{figure}

As shown in Figure \ref{fig:shape_equil}, for $1<\Delta<2$ and $n+1/2<\Delta<n+1$ ($n\in\mathds{N}_+$), the shape functions peak at the equilateral shape $k_1=k_2=k_3$, which degenerates with the contact interaction $(\nabla\phi)^4$. 
For $n<\Delta<n+1/2$, the shape functions still peak at the equilateral shape, but they are negatively correlated, plotted in Figure \ref{fig:shape_negequil}. 
These shapes degenerate with the orthogonal shape. 

\begin{figure}[h]
    \centering
    \includegraphics[width=0.70\linewidth]{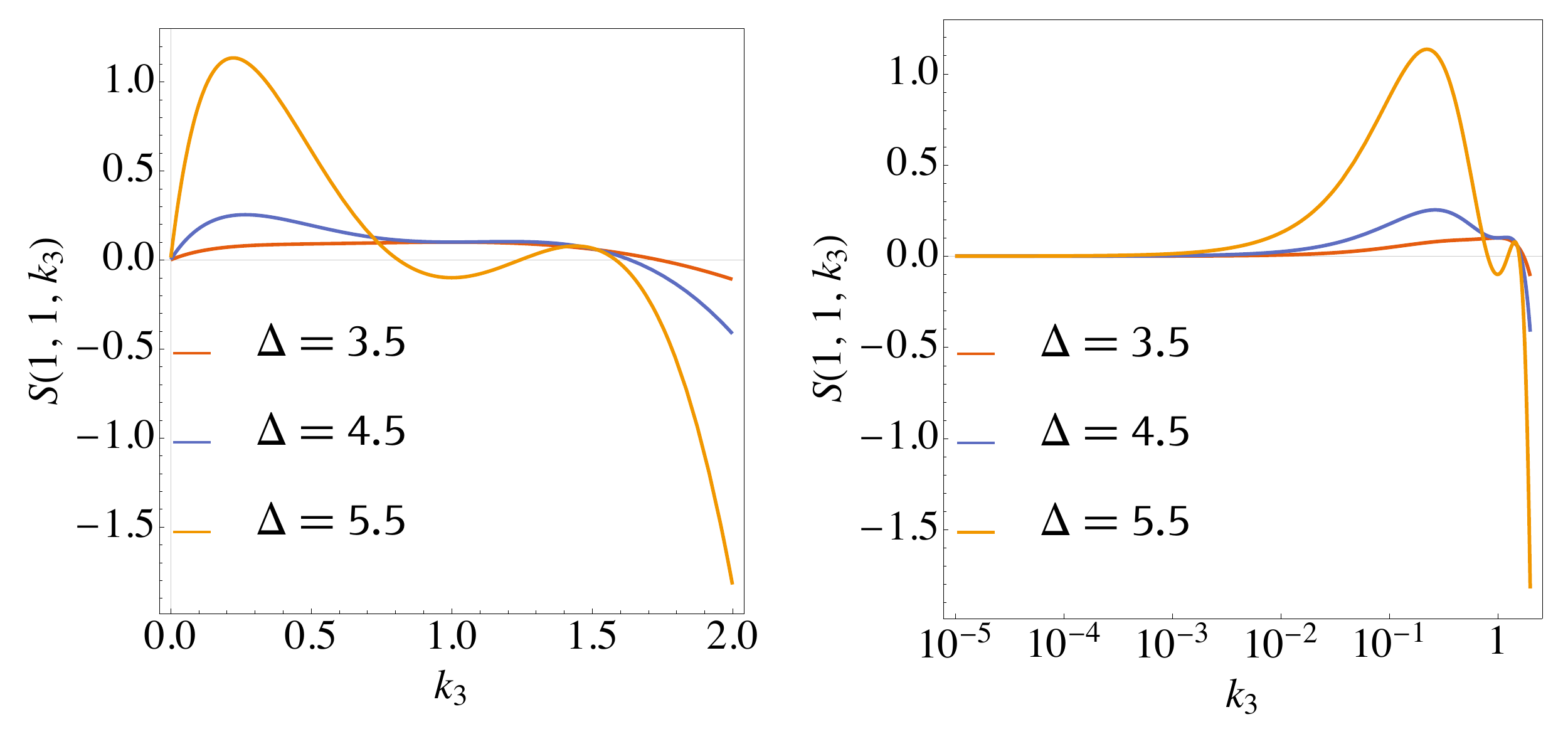}
    \caption{The shape functions when $\Delta$'s are half-integers and larger than $3$. We rescale the shape functions to make their magnitudes identical in the equilateral shape. The shape functions are oscillating throughout the range. It is the intermediate shape between Figure \ref{fig:shape_equil} and Figure \ref{fig:shape_negequil}.}
    \label{fig:shape_novel}
\end{figure}

The most special shapes are the ones for $\Delta\sim n+1/2$, when $n>3$. 
From Figure \ref{fig:shape_novel}, we can see that the shape functions for $\Delta\sim n+1/2$ ($n>3$) oscillate throughout the physical range. 
We can regard them as the intermediate shapes when the controlling parameter $\Delta$ moves continuously from shapes in Figure \ref{fig:shape_equil} to the ones in Figure \ref{fig:shape_negequil}. 
They are different from the shape functions for tree-level free massive scalar exchange, which oscillate near the squeezed limit, as in Figure \ref{fig:compare_massive_unparticle}. 
Thus this is a new kind of signal to look for. 


\newpage
\section{Conclusions and Future Directions}
\label{sec:Outlook}

In this paper, we considered a gapless sector with a large anomalous dimension that couples weakly to the curvature perturbation. 
It is perhaps the simplest example of a strongly coupled sector during inflation, in the form of unparticles. 
We have calculated the four-point late-time correlator of conformally coupled scalars in de Sitter analytically using Mellin-Barnes integration and derived its governing differential equations using twisted cohomology. 
From the four-point function in de Sitter, we obtained primordial trispectra and bispectra using weight-shifting operators. 
Unlike the massive exchange case, the inflationary trispectrum does not have oscillatory behavior, but rather a specific power law decay in squeezed limits, such as $\sim s^{2\Delta-3}$ when $\Delta<3/2$ and $\sim s^0$ when $\Delta\geq3/2$.

From the plots presented in Section \ref{sec:comments_pheno}, three typical shapes can be identified. 
In general, when $\Delta<2$ and $n+1/2<\Delta<n+1$, where $n$ is a positive integer, the shapes are similar to the equilateral shape (e.g. in \cite{Alishahiha:2004eh,Chen:2006nt,Langlois:2008wt,Arkani-Hamed:2003juy,Senatore:2004rj}). 
When $n<\Delta<n+1/2$, the shapes are similar to the orthogonal shape \cite{Senatore:2009gt}. 
The new shapes given by $\Delta\sim n+1/2$ are somewhat ``in between" the equilateral shapes and the orthogonal shapes, in which the equilateral limit is not a maximum but rather a saddle point. 
It would be interesting to see which values of $\Delta$ have small ``cosine" compared to the standard shapes of non-Gaussianity \cite{Babich:2004gb}. 
Notice that all these shapes suffer from heavy suppression that increases with power of $\Delta$.

We have just scratched the surface of a rich theoretical landscape of strongly coupled sectors in inflation, opening new avenues for future exploration. 
We list some below: 
\begin{itemize}
    \item In this work, we study the gapless scenario. 
    As mentioned in Section \ref{sec:holography}, gapped sectors could have new shapes of non-Gaussianities, especially for gaps close to the Hubble scale. 
    What exactly happens there? 
    Understanding the gapped scenarios would give us a more complete picture on how in general the strongly coupled sector behaves. Even more interestingly, we could wonder about a composite scenario, where all fluctuations, including the curvature, are generated from more fundamental fields.
    \item During our discussions, conformal symmetries play an important role for properties of unparticles. 
    In order to enhance non-Gaussianities, it would be important to incorporate the effects of a small speed of sound, as discussed in \cite{Pimentel:2022fsc}. How would the resulting shapes change? 
    \item Some recent developments in combinatorial/geometric structures contained in cosmological correlators appeared in \cite{Arkani-Hamed:2017fdk,Arkani-Hamed:2023bsv,Arkani-Hamed:2023kig,Baumann:2024mvm,Hang:2024xas,Fevola:2024nzj,Arkani-Hamed:2024jbp}. 
    In this work we have seen that the non-integer part of the unparticle scaling dimension can also act as the ``twist parameter." 
    Can we systematically develop a theory of differential equations for tree-level unparticle diagrams? 
    \item We discussed the unparticles as a hypothetical field with certain symmetries. 
    Can we develop some practical models for these unparticles in the strongly coupled sector? 
    For example, the four-dimensional Banks-Zaks model \cite{Banks:1981nn} becomes a gapless unparticle model in the IR \cite{Georgi:2008pq}, and is asymptotically free in the UV. 
    Ideally, the models should contain parameters which could characterize the gapped/gapless phases. 
    A recent model building on this scenario is discussed in \cite{Yang:2025apy}. 
    \item Finally, it would be interesting to put bounds on unparticle physics during inflation using observational data, for example, from the Cosmic Microwave Background (CMB) or Large Scale Structure (LSS).  
\end{itemize}
The microscopic nature of inflation remains mysterious. 
It is important to investigate different dynamical mechanisms for its origin and their phenomenological consequences. 
From a formal perspective, the study of more general scenarios for cosmological correlators will improve our understanding of QFT in cosmological backgrounds. 
With the development of new theoretical tools and upcoming cosmological measurements, the time is ripe to investigate the beginnings of our universe.

\section*{Acknowledgements}
Thanks to Nima Arkani-Hamed, Paolo Benincasa, Bruno Bucciotti, Craig Clark, Nathaniel Craig, Thomas Dumitrescu, Yikun Jiang, Austin Joyce, Juan Maldacena, Jiaxin Qiao, Augusto Sagnotti, Raman Sundrum, Tom Westerdijk, and Matias Zaldarriaga for useful discussions. 
Thanks to Enrico Pajer, Charlotte Sleight, Massimo Taronna and Zhenbin Yang for comments. 
GLP presented some of these results at the workshop ``What is Particle Theory?" in KITP, Santa Barbara, as well as in seminars at UCLA and IAS Princeton, and he is grateful for many interesting discussions and feedback from the participants. 
GLP thanks the KITP, as well as IAS Princeton for their hospitality and support.
CY thanks the Institute for Advanced Study, Tsinghua---in particular, Yixu Wang and Zhenbin Yang---for hosting two visits. 

GLP and CY are supported by Scuola Normale and by INFN (IS GSS-Pi). This research was supported in part by grant NSF PHY-2309135 to the Kavli Institute for Theoretical Physics (KITP). 
The research of GLP and CY is moreover supported by the ERC (NOTIMEFORCOSMO, 101126304). 
Views and opinions expressed are, however, those of the author(s) only and do not necessarily reflect those of the European Union or the European Research Council Executive Agency. 
Neither the European Union nor the granting authority can be held responsible for them. 
GLP is further supported by a Rita-Levi Montalcini Fellowship from the Italian Ministry of Universities and Research (MUR), as well as under contract 20223ANFHR (PRIN2022).

\appendix
\section{Basic Properties of the Gamma Function $\Gamma(z)$, Hypergeometric Function ${}_2F_1$ and Appell Function $F_1$}
\label{app:special_functions}

Here we present some properties of the gamma function $\Gamma(z)$, hypergeometric function ${}_2F_1$ and Appell function $F_1$ that we used in our calculations. 
A more comprehensive introduction to generalized hypergeometric functions can be found in \cite{Slater:1966fan}. 

The gamma function is an extended definition of factorial, which relates to factorial by the shift relation 
\begin{equation}
    \Gamma(n) \equiv (n-1) \Gamma(n-1) = (n-1)!\ .
    \label{def:gamma-shift}
\end{equation}
The gamma function defined on the complex plane is 
\begin{equation}
    \Gamma(z) \equiv \int_0^\infty \dint t\ e^{-t} t^{z-1} \quad \text{for }\text{Re } z>0. 
\end{equation}
When $\text{Re } z<0$, we can use the shift relation \eqref{def:gamma-shift} to analytically continue the gamma function to all non-integer values. 
The non-positive real integer $z$'s cannot be analytic continued because the shift relation is not well-defined, e.g. $\Gamma(0) = \Gamma(1)/0$. 

The hypergeometric function ${}_2F_1$ is defined by the Gauss series 
\begin{equation}
    {}_2F_1 (a,b;c;z) \equiv \sum_{m=0}^{\infty} \frac{(a)_m (b)_m}{(c)_m} \frac{z^m}{m!}, 
\end{equation}
where $(a)_m\equiv\Gamma(a+m)/\Gamma(a)$ is called Pochhammer symbol.
From this definition, it is obvious that ${}_2F_1$ is invariant under the exchange of $a$ and $b$. 
Based on the poles of the gamma function, the hypergeometric function ${}_2F_1$ is not defined when $c$ is a non-positive real integer, e.g. $0,-1,-2,\dots$ 

Some variable transformation rules for ${}_2F_1$ are 
\begin{align}
    {}_2F_1 (a,b;c;z) &= (1-z)^{-a} {}_2F_1 \big(a,c-b;c;\frac{z}{z-1}\big) = (1-z)^{-b} {}_2F_1 \big(c-a,b;c;\frac{z}{z-1}\big) \nonumber \\
    &= (1-z)^{c-a-b} {}_2F_1 (c-a,c-b;c;z). 
    \label{def:2F1_variable_trans}
\end{align}
We can use them to simplify or symmetrize the expressions. 

The Mellin-Barnes integral representation (contour integral representation) for ${}_2F_1$ is 
\begin{equation}
    {}_2F_1(a,b;c;z) = \frac{\Gamma(c)}{\Gamma(a)\Gamma(b)} \int_{-i\infty}^{+i\infty} \frac{\dint s}{2\pi i} \frac{\Gamma(a+s) \Gamma(b+s)}{\Gamma(c+s)} \Gamma(-s)\ (-z)^s. 
    \label{def:standard_2F1}
\end{equation}
Another well-known Mellin-Barnes representation for ${}_2F_1$ is 
\begin{equation}
    {}_2F_1(a,b;c;z) = \frac{\Gamma(c)}{\Gamma(a)\Gamma(b)\Gamma(c-a)\Gamma(c-b)} \int_{-i\infty}^{+i\infty} \frac{\dint s}{2\pi i} \Gamma(a+s) \Gamma(b+s) \Gamma(c-a-b-s) \Gamma(-s)\ \big(1-z\big)^s. 
    \label{def:equiv_2F1}
\end{equation}

The Appell function is defined when each variable by itself satisfies a partial differential equation that resembles the hypergeometric differential equation, but the series cannot be expressed as a product of two hypergeometric functions. 
In this paper, we used the Appell $F_1$ function, which is a two-variable generalized hypergeometric function defined as 
\begin{equation}
    F_1 (\alpha;\beta,\beta';\gamma;x,y) = \sum_{m=0}^{\infty}\sum_{n=0}^{\infty} \frac{(\alpha)_{m+n}(\beta)_m(\beta')_n}{(\gamma)_{m+n}} \frac{x^m y^n}{m!n!}. 
\end{equation}
It is also obvious from this definition that the Appell $F_1$ function is invariant when we exchange $\beta\leftrightarrow\beta'$ and $x \leftrightarrow y$ at the same time. 
Similar to the hypergeometric function, the Appell $F_1$ function is not defined when $\gamma$ is a non-positive real integer. 

Some variable transformation rules for $F_1$ are 
\begin{align}
    F_1 (\alpha; \beta, \beta'; \gamma; x, y) &= (1-x)^{-\alpha} F_1 \big(\alpha; \gamma-\beta-\beta', \beta'; \gamma; \frac{x}{x-1}, \frac{y-x}{1-x}\big), \\
    &= (1-x)^{-\beta} (1-y)^{-\beta'} F_1 \big(\gamma-\alpha; \beta, \beta'; \gamma; \frac{x}{x-1}, \frac{y}{y-1}\big). 
    \label{def:F1_variable_trans}
\end{align}

The standard Mellin-Barnes representation for Appell $F_1$ function is 
\begin{align}
    F_1 (\alpha;\beta,\beta';\gamma;x,y) &= \frac{\Gamma(\gamma)}{\Gamma(\alpha)\Gamma(\beta)\Gamma(\beta')} \int_{-i\infty}^{+i\infty} \int_{-i\infty}^{+i\infty} \frac{\dint s \dint t}{(2\pi i)^2} \frac{\Gamma(\alpha+s+t)\Gamma(\beta+s)\Gamma(\beta'+t)}{\Gamma(\gamma+s+t)} \nonumber \\
    &\times \Gamma(-s) \Gamma(-t) (-x)^s (-y)^t. 
    \label{def:standard_F1}
\end{align}
Notice that we can use the equivalence between \eqref{def:standard_2F1} and \eqref{def:equiv_2F1} to find an equivalent form for Appell $F_1$, which is 
\begin{align}
    &F_1 (\alpha;\beta,\beta';\gamma;x,y) = \frac{\Gamma(\gamma)}{\Gamma(\alpha)\Gamma(\beta)\Gamma(\beta')\Gamma(\gamma-\alpha)\Gamma(\gamma-\beta-\beta')} \int_{-i\infty}^{+i\infty} \int_{-i\infty}^{+i\infty} \frac{\dint s \dint t}{(2\pi i)^2} \Gamma(\alpha+s+t) \nonumber \\
    &\times \Gamma(\beta+s) \Gamma(\beta'+t) \Gamma(\gamma-\alpha-\beta-\beta'-s-t) \Gamma(-s) \Gamma(-t) (1-x)^{s} (1-y)^{t}. 
    \label{def:equiv_F1}
\end{align}

From \cite{Slater:1966fan}, we know that Appell $F_1$ function relates to hypergeometric function ${}_2F_1$ through a Mellin-Barnes integration: 
\begin{align}
    &\int^{+i\infty}_{-i\infty} \frac{\dint s}{2\pi i}\ \frac{\Gamma(b-s)\Gamma(g+s)\Gamma(h-s)}{\Gamma(d-s)} {}_2F_1 \big(b-s, e, d-s; x\big) y^{-s} \nonumber \\
    =& y^g \frac{\Gamma(b+g)\Gamma(h+g)}{\Gamma(d+g)} F_1 \big(b+g; e, h+g; d+g; x, -y\big). 
    \label{def:inte_over_2F1_to_F1}
\end{align}
The Appell $F_1$ function can be expressed as a sum of hypergeometric functions: 
\begin{align}
    &F_1 (\alpha; \beta, \beta'; \gamma; x, y) = \sum_m \frac{x^m}{m!} \frac{(\alpha)_m (\beta)_m}{(\gamma)_m} {}_2F_1 \big(\alpha+m, \beta', \gamma+m; y\big). 
    \label{def:sum_over_2F1_to_F1}
\end{align}

The Appell $F_1$ function can reduce to the hypergeometric function ${}_2F_1$ in special cases. 
For example, if $\gamma=\beta+\beta'$, the reduction gives 
\begin{equation}
    F_1 (\alpha; \beta, \beta'; \beta+\beta'; x, y) = (1-y)^{-\alpha} {}_2F_1 \big(\alpha,\beta,\beta+\beta';\frac{x-y}{1-y}\big). 
    \label{def:F1_equal_coeff_reduction}
\end{equation}
If $x=y$, the Appell $F_1$ function reduces as 
\begin{align}
    F_1 (\alpha; \beta, \beta'; \gamma; x, x) &= (1-x)^{\gamma-\alpha-\beta-\beta'} {}_2F_1 (\gamma-\alpha,\gamma-\beta-\beta',\gamma;x) \\
    &= {}_2F_1 (\alpha, \beta+\beta', \gamma; x). 
\end{align}
Finally, the Appell $F_1$ function trivially reduces to ${}_2F_1$ if one of the two variables is zero: 
\begin{equation}
    F_1 (\alpha; \beta, \beta'; \gamma; x, 0) = {}_2F_1 (\alpha,\beta,\gamma,x). 
\end{equation}

\section{Details of the Mellin-Barnes Integration}
\label{app:MB}
For simplicity, we can denote $\epsilon\equiv\Delta-2$. 
Thus the target integrals are 
\begin{align}
    I_1 &= \int^\infty_0 \dint x_1 \int^\infty_0 \dint x_2  \int^\infty_1 \dint t \ (x_1 x_2)^{-1-\epsilon} (t^2-1)^{\epsilon}\ \frac{1}{x_1+t+\frac{1}{u}} \frac{1}{x_1+x_2+\frac{1}{u}+\frac{1}{v}}, \\
    I_2 &= \int^\infty_0 \dint x_1 \int^\infty_0 \dint x_2  \int^\infty_1 \dint t \ (x_1 x_2)^{-1-\epsilon} (t^2-1)^{\epsilon}\ \frac{1}{x_1+t+\frac{1}{u}} \frac{1}{x_2+t+\frac{1}{v}}. 
\end{align}
We will use Mellin-Barnes integration to perform these integrals: \\
For $I_1$, we have 
\begin{equation}
    \begin{aligned}
    &\frac{1}{x_1+x_2+\frac{1}{u}+\frac{1}{v}} = \frac{1}{\Gamma(1)} \int^{c_1+i\infty}_{c_1-i\infty} \frac{\dint z_1}{2\pi i}\ \Gamma(1+z_1) \Gamma(-z_1) (x_1+\frac{1}{u})^{z_1} (x_2+\frac{1}{v})^{-1-z_1} \\
    &\frac{1}{x_1+\frac{1}{u}+t+1} = \frac{1}{\Gamma(1)} \int^{c_2+i\infty}_{c_2-i\infty} \frac{\dint z_2}{2\pi i}\ \Gamma(1+z_2) \Gamma(-z_2) (1+x_1+\frac{1}{u})^{-1-z_2}  t^{z_2}, 
\end{aligned}
\end{equation}
and we need to further expand 
\begin{align}
    \frac{1}{(1+x_1+\frac{1}{u})^{1+z_2}} = \frac{1}{\Gamma(1+z_2)} \int^{c_3+i\infty}_{c_3-i\infty} \frac{\dint z_3}{2\pi i}\ \Gamma(z_2+z_3+1) \Gamma(-z_3) (x_1+\frac{1}{u})^{z_3}. 
\end{align}
Now it is time to perform the real integrals: 
\begin{equation}
    \begin{aligned}
        &\int^\infty_0 \dint x_1\ x_1^{-1-\epsilon} (x_1+\frac{1}{u})^{z_1+z_3} = \Big(\frac{1}{u}\Big)^{z_1+z_3-\epsilon} \frac{\Gamma(-\epsilon)\Gamma(-z_1-z_3+\epsilon)}{\Gamma(-z_1-z_3)}, \\
        &\int^\infty_0 \dint x_2\ x_2^{-1-\epsilon} (x_2+\frac{1}{v})^{-1-z_1} = \Big(\frac{1}{v}\Big)^{-z_1-1-\epsilon} \frac{\Gamma(-\epsilon)\Gamma(z_1+1+\epsilon)}{\Gamma(1+z_1)}, \\
        &\int^\infty_0 \dint t\ t^{z_2+\epsilon} (t+2)^{\epsilon} = 2^{1+2\epsilon+z_2} \frac{\Gamma(z_2+1+\epsilon)\Gamma(-z_2-1-2\epsilon)}{\Gamma(-\epsilon)}, 
    \end{aligned}
\end{equation}
Collecting the results above, the Mellin-Barnes representation of $I_1$ will be 
\begin{align}
    I_1 &= 2^{2\epsilon+1} u^{\epsilon} v^{\epsilon+1} \Gamma(-\epsilon) \int_{C_1}\int_{C_2}\int_{C_3} \frac{\dint z_1 \dint z_2 \dint z_3}{(2\pi i)^3}\ \Gamma(z_1+1+\epsilon) \Gamma(z_2+1+\epsilon) \Gamma(-z_2-1-2\epsilon) \nonumber \\
    &\times \Gamma(z_2+z_3+1) \frac{\Gamma(-z_1-z_3+\epsilon)}{\Gamma(-z_1-z_3)} \Gamma(-z_1) \Gamma(-z_2) \Gamma(-z_3) \big(\frac{u}{v}\big)^{-z_1} \big(\frac{1}{2}\big)^{-z_2} u^{-z_3} \\
    &= \frac{\Gamma(-\epsilon) \Gamma(1+\epsilon)}{2} \frac{\pi}{\sin(2\pi\epsilon)} \Bigg(\Big(\frac{2uv}{u+v}\Big)^{2\epsilon+2} \frac{1}{1+\epsilon} F_1 \Big(1; 1+\epsilon, 1+\epsilon; 2+\epsilon; \frac{u(1-v)}{u+v}, \frac{u(1+v)}{u+v}\Big) \nonumber \\
    &+ \Big(\frac{2uv}{u+v}\Big) \frac{1}{\epsilon} F_1 \Big(1; -\epsilon, -\epsilon; 1-\epsilon; \frac{u(1-v)}{u+v}, \frac{u(1+v)}{u+v}\Big)\Bigg). 
\end{align}
During the computation, we perform this integration in the order $C_3\Rightarrow C_2\Rightarrow C_1$, while using \eqref{def:standard_2F1}, \eqref{def:inte_over_2F1_to_F1}, \eqref{def:sum_over_2F1_to_F1}, \eqref{def:F1_equal_coeff_reduction} and the transformation of variables. 

For $I_2$, the $x_1$, $x_2$ integrals can be directly performed: 
\begin{align}
    &\int^\infty_0 \dint x_1\ x_1^{-1-\epsilon}\ \frac{1}{x_1+t+\frac{1}{u}} = \Gamma(-\epsilon)\Gamma(1+\epsilon) \Big(\frac{1}{t+\frac{1}{u}}\Big)^{1+\epsilon}, \\
    &\int^\infty_0 \dint x_2\ x_2^{-1-\epsilon}\ \frac{1}{x_2+t+\frac{1}{v}} = \Gamma(-\epsilon)\Gamma(1+\epsilon) \Big(\frac{1}{t+\frac{1}{v}}\Big)^{1+\epsilon}. 
\end{align}
Now we can expand the integrand as 
\begin{equation}
    \begin{aligned}
        &\frac{1}{(t+\frac{1}{u}+1)^{\epsilon+1}} = \frac{1}{\Gamma(\epsilon+1)}\frac{1}{2\pi i} \int^{+i\infty}_{-i\infty} \dint z_1\ \Gamma(z_1+\epsilon+1)\Gamma(-z_1)t^{-\epsilon - 1 - z_1} (\frac{1}{u}+1)^{z_1}, \\
        &\frac{1}{(t+\frac{1}{v}+1)^{\epsilon+1}} = \frac{1}{\Gamma(\epsilon+1)}\frac{1}{2\pi i} \int^{+i\infty}_{-i\infty} \dint z_2\ \Gamma(z_2+\epsilon+1)\Gamma(-z_2)t^{-\epsilon - 1 - z_2} (\frac{1}{v}+1)^{z_2}. 
    \end{aligned}
\end{equation}
Using \eqref{def:equiv_F1}, we can write out the result for $I_2$: 
\begin{align}
    I_2 = \frac{\Gamma^2(-\epsilon)\Gamma^2(1+\epsilon)}{2(1+\epsilon)}\  F_1\Big(1;1+\epsilon,1+\epsilon;2+\epsilon;\frac{u-1}{2u},\frac{v-1}{2v}\Big). 
\end{align}

Here we have also computed the general analytic form for the integral \eqref{def:general_I++}: 
\begin{align}
    \psi &= 2^{1+2\epsilon_2} u^{-1-\epsilon_1} v^{-\epsilon_1} \frac{\Gamma^2(1+\epsilon_1)}{\Gamma(-\epsilon_2)} \int_{C_1}\int_{C_2}\int_{C_3} \frac{\dint z_1 \dint z_2 \dint z_3}{(2\pi i)^3}\ \Gamma(z_1-\epsilon_1) \Gamma(z_2+1+\epsilon_2) \Gamma(-z_2-1-2\epsilon_2) \nonumber \\
    &\times \Gamma(z_2+z_3+1) \frac{\Gamma(-z_1-z_3-1-\epsilon_1)}{\Gamma(-z_1-z_3)} \Gamma(-z_1) \Gamma(-z_2) \Gamma(-z_3) \big(\frac{u}{v}\big)^{-z_1} \big(\frac{1}{2}\big)^{-z_2} u^{-z_3} \\
    &= \Gamma^2(1+\epsilon_1)\frac{\Gamma(-\epsilon_1)}{\Gamma(-\epsilon_2)} 2^{1+2\epsilon_2} \left(\frac{1+u}{u}\right)^{\epsilon_1} \left(\frac{u+v}{uv}\right)^{\epsilon_1} \frac{\pi}{\sin(2\pi\epsilon_2)} \sum_{k=0}^{\infty} \frac{1}{k!}\left(\frac{u(1-v)}{u+v}\right)^k (1+\epsilon_1)_k (-\epsilon_1)_k \nonumber \\
    &\times \Bigg(\frac{\Gamma(-2\epsilon_1)\Gamma(1+\epsilon_2)}{\Gamma(2+2\epsilon_2)} \frac{1}{\Gamma(1-\epsilon_1+k)} {}_3F_2 \Big(1,-2\epsilon_1,1+\epsilon_2;2+2\epsilon_2,1-\epsilon_1+k;\frac{2u}{1+u}\Big) \nonumber \\
    &- \left(\frac{u+1}{2u}\right)^{1+2\epsilon_2} \frac{\Gamma(-1-2\epsilon_1-2\epsilon_2)\Gamma(-\epsilon_2)}{\Gamma(-\epsilon_1-2\epsilon_2+k)} {}_2F_1\Big(-\epsilon_2,-1-2\epsilon_1-2\epsilon_2;-\epsilon_1-2\epsilon_2+k;\frac{2u}{u+1}\Big)\Bigg). 
\end{align}
It is easy to check that if we plug in the twist parameters for unparticles in four-dimensional de Sitter space, the generalized hypergeometric function ${}_3F_2$ reduces to an ordinary hypergeometric function ${}_2F_1$, which allows us to perform the summation. 
\section{Shape of Bispectra with Spinning Exchange}
\label{app:SE}

In this section, we discuss the phenomenological aspect of bispectra with spinning-unparticle exchanges, mainly focused on the stress tensor of scalar unparticles. 

Taking the soft limit $k_4\rightarrow0$ will enormously reduce the number of kinematic variables. 
Based on the momentum conservation $\delta^{(3)}(\vec{k}_1+\vec{k}_2+\vec{k}_3)$, we will find 
\begin{equation}
    \begin{aligned}
        s\rightarrow k_3,\ v \rightarrow 1,\ \beta \rightarrow k_3,\ \tau \rightarrow (\vec{k}_1-\vec{k}_2)\cdot \vec{k}_3 = k_2^2-k_1^2\ \Rightarrow\ \hat{T} \rightarrow 0,\ \hat{L}^2 \rightarrow 0. 
    \end{aligned}
\end{equation}
The fact that $\hat{T}$ and $\hat{L}^2$ vanish leads to the disappearance of the polarization sum of the transverse modes, $\bar{\Pi}_m(\tau,\alpha,\beta)$. 
Therefore the induced bispectra only receive contributions from the longitudinal modes. 
It is worthy of pointing out that now the bispectra depend on only one more external variable, $\alpha$, which is the difference between $k_1$ and $k_2$. 

\begin{figure}[h]
    \centering
    \includegraphics[width=0.75\linewidth]{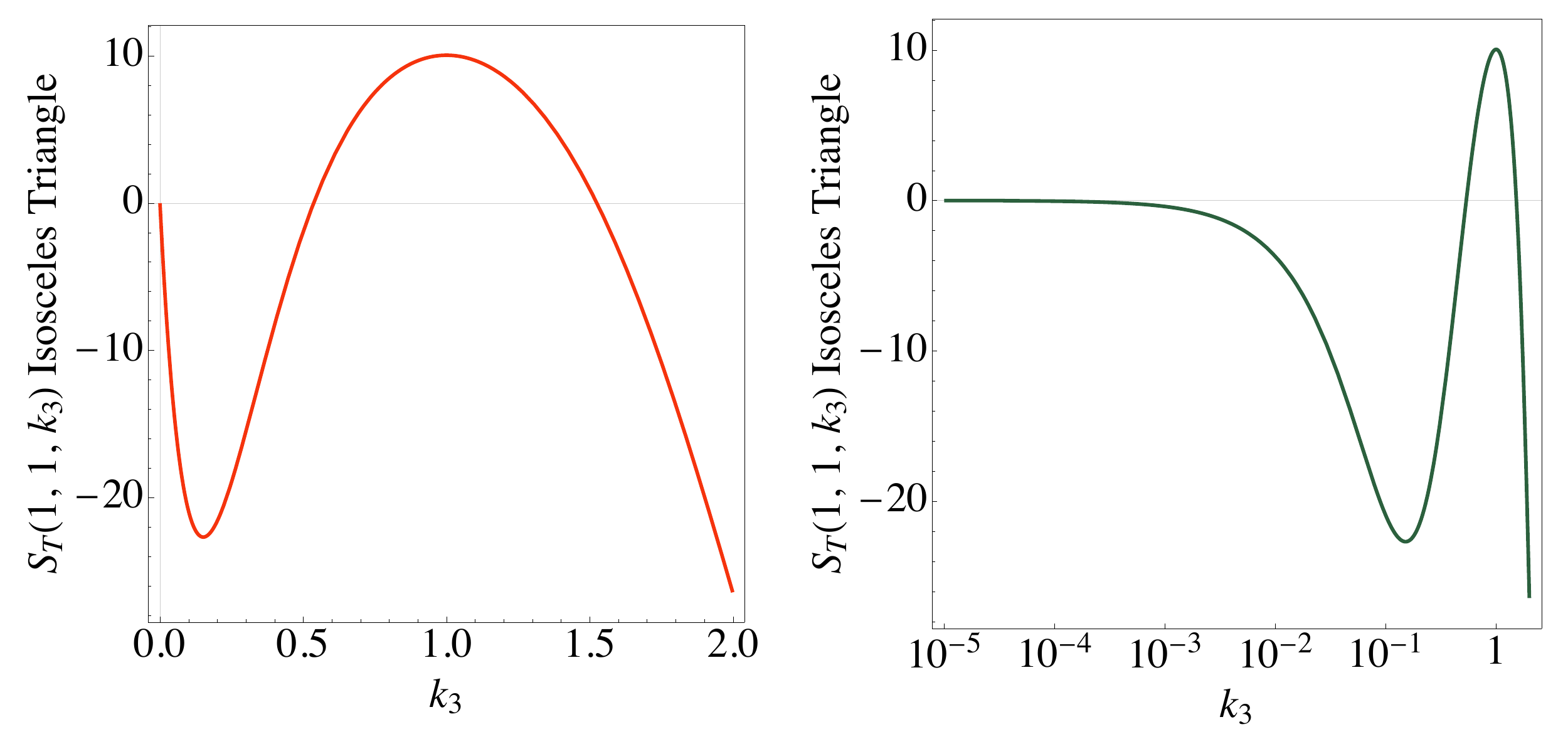}
    \caption{The isosceles triangular shape function for unparticle-stress-tensor exchange. The left panel shows the behavior near the squeezed limit. The curve in the right panel peaks at the equilateral shape, while having a negative extremum in the folded limit. Here we take $\epsilon=2$ for simplicity. }
    \label{fig:exchange_stresstensor}
\end{figure}

For concreteness, we only consider the four-point function of inflatons interacting with the spin-2 unparticle stress tensor through a minimal number of derivatives, to induce the three-point function. 
We will study the interaction vertex where the stress tensor of the unparticle couples to a higher-derivative tensor of $\phi$, defined as 
\begin{align}
    T^{\phi}_{\mu\nu} \equiv \nabla_\mu \nabla_\nu \nabla_\alpha\phi_{k_1}\nabla^\alpha\phi_{k_2} - 2\nabla_\mu\nabla_\alpha\phi_{k_1} \nabla_\nu\nabla^\alpha\phi_{k_2} - 3\nabla_\mu\phi_{k_1}\nabla_\nu\phi_{k_2} + (k_1\leftrightarrow k_2). 
\end{align}
The induced bispectrum for spin-2 stress tensor exchange will be \cite{Arkani-Hamed:2018kmz}: 
\begin{align}
    &\mathcal{B}_{\phi}^{(S=2)}(k_1,k_2,k_3) = \frac{\epsilon}{2}k_3^3 P_2\left(\alpha/k_3\right) U_{12}^{2,0} \left(\Delta_u(\Delta_u-2)\ F_{\Delta=4}(u,1)\right)\ +\ \text{permutations}, \\
    &U_{12}^{2,0} \equiv U_{12}^{1,0}(\cdot) + \left(\frac{\alpha^2/s^2}{P_2(\alpha/s)} - \frac{1+u^2}{2u^2}\right)(\cdot). 
\end{align}
The isosceles shape (where we take $k_1=k_2=1$) is shown in Figure \ref{fig:exchange_stresstensor}. 
This shape function has two extrema, one of which is in the equilateral shape, while the true extremum occurs in the folded limit. 

\clearpage
\addcontentsline{toc}{section}{References}
\bibliographystyle{utphys}
{\linespread{1.075}
\bibliography{Refs}
}

\end{document}